\documentclass[aps,preprint,onecolumn,amsmath,amssymb,superscriptaddress]{revtex4-1}
\usepackage{amssymb}
\usepackage{graphicx}
\usepackage{dcolumn}
\usepackage{bm}
\usepackage{amsmath}
\usepackage{color}
\usepackage{amsfonts}%
\usepackage{multibib}
\usepackage[left]{lineno}
\begin{document}
\title{Engineering long spin coherence times of spin-orbit systems}
\author{T. Kobayashi} 
\email{kobayashi20131124@gmail.com}
\affiliation{Centre for Quantum Computation and Communication Technology, School of Physics, University of New South Wales Sydney, NSW 2052, Australia}
\affiliation{Department of Physics, Tohoku University, Sendai 980-8578, Japan}
\author{J. Salfi} 
\affiliation{Centre for Quantum Computation and Communication Technology, School of Physics, University of New South Wales Sydney, NSW 2052, Australia}
\author{J. van der Heijden} 
\affiliation{Centre for Quantum Computation and Communication Technology, School of Physics, University of New South Wales Sydney, NSW 2052, Australia}
\author{C. Chua} 
\affiliation{Centre for Quantum Computation and Communication Technology, School of Physics, University of New South Wales Sydney, NSW 2052, Australia}
\author{M. G. House} 
\affiliation{Centre for Quantum Computation and Communication Technology, School of Physics, University of New South Wales Sydney, NSW 2052, Australia}
\author{D. Culcer} 
\affiliation{School of Physics, University of New South Wales Sydney, NSW 2052, Australia}
\affiliation{Australian Research Council Centre of Excellence in Low-Energy Electronics Technologies, The University of New South Wales, Sydney 2052, Australia}
\author{W. D. Hutchison} 
\affiliation{School of Physical, Environmental and Mathematical Sciences, The University of New South Wales Canberra, Canberra, ACT 2600, Australia}
\author{B. C. Johnson} 
\affiliation{Centre for Quantum Computation and Communication Technology, School of Physics, University of Melbourne, VIC 3010, Australia}
\author{J. C. McCallum} 
\affiliation{Centre for Quantum Computation and Communication Technology, School of Physics, University of Melbourne, VIC 3010, Australia}
\author{H. Riemann} 
\affiliation{Leibniz-Institut f\"{u}r Kristallz\"{u}chtung, 12489 Berlin, Germany}
\author{N. V. Abrosimov} 
\affiliation{Leibniz-Institut f\"{u}r Kristallz\"{u}chtung, 12489 Berlin, Germany}
\author{P. Becker} 
\affiliation{PTB Braunschweig, 38116 Braunschweig, Germany}
\author{H.-J. Pohl} 
\affiliation{VITCON Projectconsult GmbH, 07745 Jena, Germany}
\author{M. Y. Simmons} 
\affiliation{Centre for Quantum Computation and Communication Technology, School of Physics, University of New South Wales Sydney, NSW 2052, Australia}
\author{S. Rogge} 
\email{s.rogge@unsw.edu.au}
\affiliation{Centre for Quantum Computation and Communication Technology, School of Physics, University of New South Wales Sydney, NSW 2052, Australia}
\date{\today}

\begin{abstract}
Spin-orbit coupling fundamentally alters spin qubits, opening pathways to improve the scalability of quantum computers via long distance coupling mediated by electric fields, photons, or phonons. 
It also allows for new engineered hybrid and topological quantum systems. 
However, spin qubits with intrinsic spin-orbit coupling are not yet viable for quantum technologies due to their short ($\sim1~\mu$s) coherence times $T_2$, while qubits with long $T_2$ have weak spin-orbit coupling making qubit coupling short-ranged and challenging for scale-up. 
Here we show that an intrinsic spin-orbit coupled ``generalised spin'' with total angular momentum $J=\tfrac{3}{2}$, which is defined by holes bound to boron dopant atoms in strained $^{28}$Si, has $T_2$ rivalling the electron spins of donors and quantum dots in $^{28}$Si. 
Using pulsed electron paramagnetic resonance, we obtain 0.9 ms Hahn-echo and 9 ms dynamical decoupling $T_2$ times, where strain plays a key role to reduce spin-lattice relaxation and the longitudinal electric coupling responsible for decoherence induced by electric field noise. 
Our analysis shows that transverse electric dipole can be exploited for electric manipulation and qubit coupling while maintaining a weak longitudinal coupling, a feature of $J=\tfrac{3}{2}$ atomic systems with a strain engineered quadrupole degree of freedom.
These results establish single-atom hole spins in silicon with quantised total angular momentum, not spin, as a highly coherent platform with tuneable intrinsic spin-orbit coupling advantageous to build artificial quantum systems and couple qubits over long distances.
\end{abstract}

\maketitle






Spin-orbit coupling fundamentally alters spin qubits, opening pathways to improve the scalability of quantum computers via long distance coupling mediated by electric fields, photons, or phonons \cite{Ladd10Nature,Xiang13RMP}. 
It also allows for new engineered hybrid and topological quantum systems \cite{Georgescu14RMP,Kato04Science,Sau10PRL}. 
However, spin qubits with intrinsic spin-orbit coupling are not yet viable for quantum technologies due to their short ($\sim1~\mu$s) coherence times $T_2$ \cite{Nadj-Perge10Nature,Song11EPL,Higginbotham14NL,Maurand16NComm}, while qubits with long $T_2$ have weak spin-orbit coupling \cite{Tyryshkin12NMat,Muhonen14NNano,Veldhorst14NatureNano} making qubit coupling short-ranged and challenging for scale-up. 
Here we show that an intrinsic spin-orbit coupled ``generalised spin'' with total angular momentum $J=\tfrac{3}{2}$, which is defined by holes bound to boron dopant atoms in strained $^{28}$Si, has $T_2$ rivalling the electron spins of donors and quantum dots in $^{28}$Si \cite{Tyryshkin12NMat,Muhonen14NNano,Veldhorst14NatureNano}. 
Using pulsed electron paramagnetic resonance (EPR), we obtain 0.9 ms Hahn-echo and 9 ms dynamical decoupling $T_2$ times, where strain plays a key role to reduce spin-lattice relaxation and the longitudinal electric coupling responsible for decoherence induced by electric field noise \cite{Beaudoin16Nanotech}. 
Our analysis shows that transverse electric dipole can be exploited for electric manipulation and qubit coupling \cite{Beaudoin16Nanotech} while maintaining a weak longitudinal coupling, a feature of $J=\tfrac{3}{2}$ atomic systems with a strain engineered quadrupole degree of freedom.
These results establish single-atom hole spins in silicon with quantised total angular momentum, not spin, as a highly coherent platform with tuneable intrinsic spin-orbit coupling advantageous to build artificial quantum systems and couple qubits over long distances.

\begin{figure*}
\begin{center}
\includegraphics{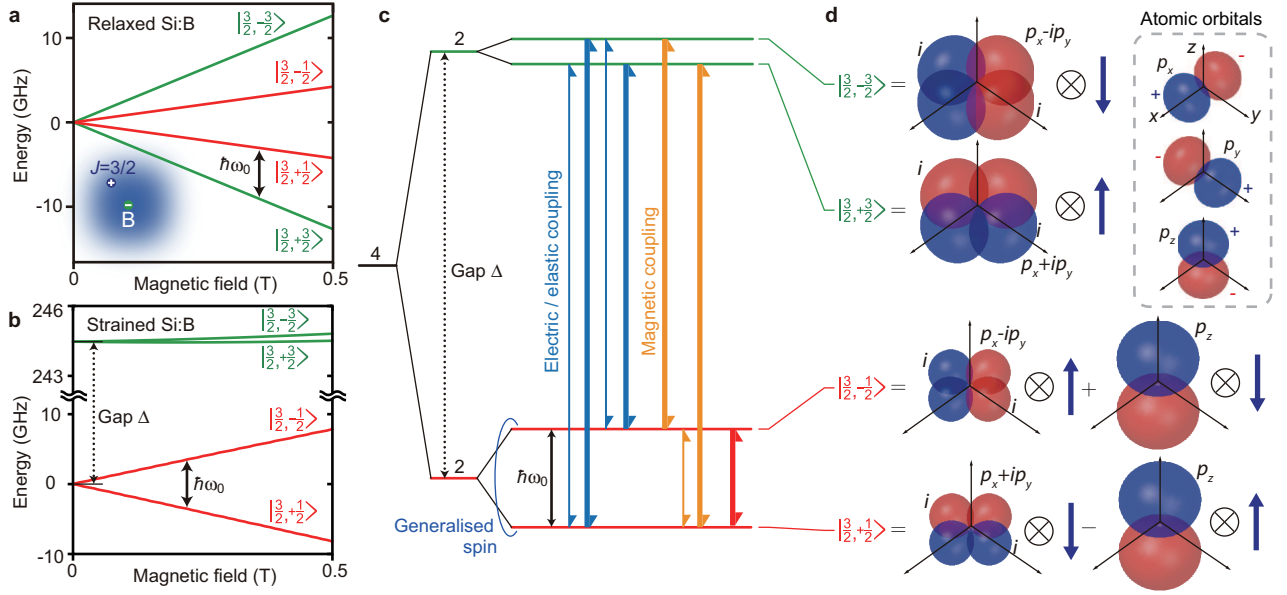}
\caption{\textbf{a,b,} Si:B hole spin levels in relaxed (\textbf{a}) and strained silicon (\textbf{b}) in an applied magnetic field, assuming a biaxial tensile strain of 0.02 \%. Black solid arrows indicate transitions addressed in this work. 
\textbf{c,} Energy level diagram and couplings for a strained (gapped) system. Spin-orbit coupling induced quadrupolar electric and elastic couplings are shown (blue arrows) as well as magnetically induced couplings (orange arrows).  
In a constant magnetic field, the quadrupolar coupling introduces an electric transition dipole (red arrow) to the $\{\left|\tfrac{3}{2},+\tfrac{1}{2}\right\rangle,\left|\tfrac{3}{2},-\tfrac{1}{2}\right\rangle\}$ generalised spin subsystem.
\textbf{d,} Schematic images of Bloch wavefunctions for $J=\tfrac{3}{2}$ hole states.
}%
\label{Concept}
\end{center}
\end{figure*}

Hole spins bound to group-III acceptors in Si have compelling properties for building qubits with spin-orbit coupling. 
These properties derive from the $\Gamma_8$ symmetry of valence-band holes where the $L=1$ angular momentum of the atomic orbitals $\left|p_{x,y,z}\right\rangle$ couples to spin $S=\tfrac{1}{2}$.
As a result, the total angular momentum $J=\tfrac{3}{2}$ is a good quantum number and the Bloch states (Fig.~\ref{Concept}d), described by projected angular momentum $m_J=\pm\tfrac{1}{2}$ (light holes) and $m_J=\pm\tfrac{3}{2}$ (heavy holes), can be written as \cite{Luttinger55PR}
\begin{eqnarray}
\left|\tfrac{3}{2},\pm\tfrac{1}{2}\right\rangle &=& \tfrac{1}{\sqrt{6}}\left( \left|p_x \right\rangle \pm i\left|p_y\right\rangle\right)\otimes\left|S_z=\mp\tfrac{1}{2}\right\rangle \mp \tfrac{2}{\sqrt{6}}\left|p_z\right\rangle\otimes\left|S_z=\pm\tfrac{1}{2}\right\rangle,\text{ and}\nonumber\\
\left|\tfrac{3}{2},\pm\tfrac{3}{2}\right\rangle &=& \tfrac{1}{\sqrt{2}}\left( \left|p_x\right\rangle \pm i\left|p_y\right\rangle\right)\otimes\left|S_z=\pm\tfrac{1}{2}\right\rangle\nonumber,
\end{eqnarray}
where $\left|S_z=\pm1/2\right\rangle$ are spin up/down.  For group-III acceptors, different types of qubits can be defined by using the low-energy $J=\tfrac{3}{2}$ manifold.  
In a perfect silicon crystal, the two lowest-energy states form a charge-like subsystem  $\{\left|\tfrac{3}{2},+\tfrac{3}{2}\right\rangle,\left|\tfrac{3}{2},+\tfrac{1}{2}\right\rangle\}$ (Fig.~\ref{Concept}a).  
Lowering the crystal symmetry by mechanical strain results in a gap $\Delta$ (Figs.~\ref{Concept}b and c), defining another qubit subsystem $\{\left|\tfrac{3}{2},+\tfrac{1}{2}\right\rangle,\left|\tfrac{3}{2},-\tfrac{1}{2}\right\rangle\}$.
This qubit is referred to as generalised spin because it is time-reversal symmetric while spin $S$ is not a good quantum number.
Coupling to electric and elastic fields for hole spins (blue arrows, Fig.~\ref{Concept}c) takes place via quadrupolar tensor operators $\hat{Q}_{ij}$ (see Supplemental material) that are represented by quadratic forms of spin-$\tfrac{3}{2}$ matrices $\hat{J}_{x,y,z}$ and have no analogue in the conduction band \cite{Winkler04PRB}.  
Combined with the $J=\tfrac{3}{2}$ Zeeman interaction (orange arrows, Fig.~\ref{Concept}c), the quadrupolar couplings endow the generalised spin with intrinsic spin-orbit coupling (red arrow, Fig.~\ref{Concept}c).
This contrasts recent work on electron quantum dot systems in silicon where spin-orbit coupling is induced with extrinsic sources such as charge degrees of freedom \cite{Kim14Nature} and micron-scale magnets \cite{Kawakami14NNano,Yoneda18NNano,Zajac18Science}. 
Recently it has been predicted that the quadrupoles allow longitudinal electric couplings responsible for qubit decoherence to be minimized while maintaining spin-orbit qubit functionality via large transverse electric coupling \cite{Salfi16PRL}.
Further advantages of acceptors in Si compared to conventional spin qubits include removal of the nuclear spin bath by $^{28}$Si purification \cite{Tyryshkin12NMat}, single-atom addressability \cite{vanderHeijden17arXiv}, confinement of spin without gate electrodes \cite{Muhonen14NNano,vanderHeijden17arXiv}, and reduced influence from charge traps at interfaces.
Nevertheless, the coherence of acceptor-bound hole spins has received little attention \cite{Song11EPL}.

\begin{figure*}
[ptb]
\begin{center}
\includegraphics{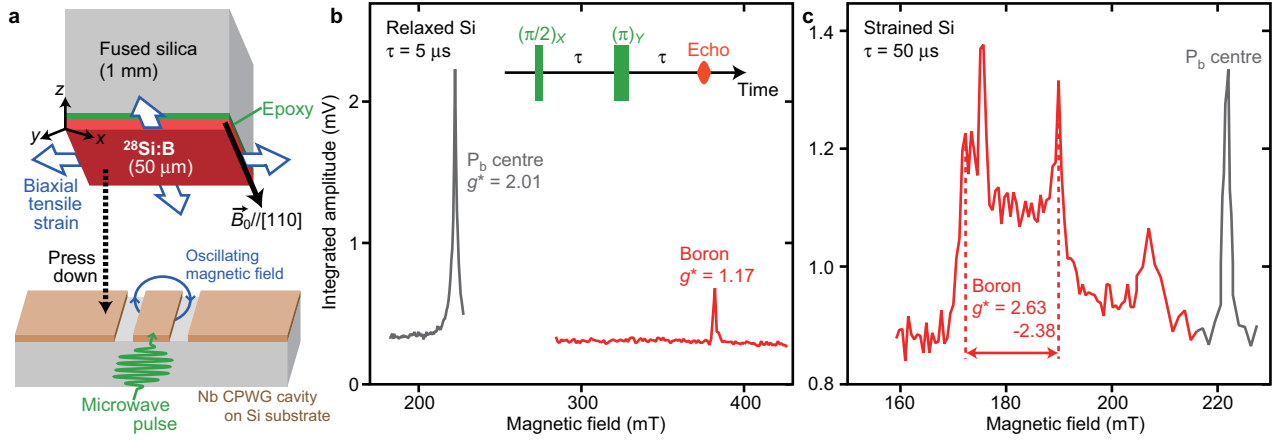}
\caption{\textbf{a,} Schematic figure of the sample and the cavity. A $50~\mu\text{m}$ thin $^{28}$Si (001) chip is bonded to a $1~\text{mm}$-thick fused-silica chip by two-component epoxy resin. A biaxial tensile strain is induced in the $^{28}$Si chip at cryogenic temperatures due to thermal expansion mismatch of Si and fused silica. The magnetic field is applied along the [110] crystal direction of the $^{28}$Si chip. To perform EPR experiments, the sample stack is pressed down to a superconducting Nb coplanar waveguide resonator. 
\textbf{b,c,} Spin-echo spectra for the mechanically relaxed (\textbf{b}) and strained (\textbf{c}) samples, measured with $\tau = 5~\mu\text{s}$ and $50~\mu\text{s}$ respectively.
}%
\label{Spectra}
\end{center}
\end{figure*}

Here we experimentally explore the effect of a strain-induced gap on the electric quadrupole and coherence of holes bound to boron acceptors in silicon 28 ($^{28}$Si:B).
To study effects of strain in coherence, we use two bulk $^{28}$Si crystal samples with and without mechanical strain.
The mechanically relaxed sample provides the energy level configuration in Fig.~\ref{Concept}a ($\Delta=0$), where we investigate the $\{\left|\tfrac{3}{2},+\tfrac{3}{2}\right\rangle,\left|\tfrac{3}{2},+\tfrac{1}{2}\right\rangle\}$ subsystem.
Another sample is subjected to biaxial tensile strain (Fig.~\ref{Spectra}a) to obtain $\Delta$ exceeding the qubit energy splitting $\hbar\omega_0$ (Figs.~\ref{Concept}b and c).
In this gapped configuration, we investigate the $\{\left|\tfrac{3}{2},+\tfrac{1}{2}\right\rangle,\left|\tfrac{3}{2},-\tfrac{1}{2}\right\rangle\}$ generalised spin-subsystem (Fig.~\ref{Concept}c), which has been predicted to have enhanced immunity to decoherence from electrical noise \cite{Salfi16PRL} and has a reduced longitudinal relaxation rate \cite{Dirksen89JPCM,Ruskov13PRB}.
For both the strained and relaxed samples we use low-temperature (base temperature $T_b\approx25~\text{mK}$) pulsed EPR (see Supplemental material) to measure the qubit $T_2$ with and without dynamical decoupling, and the longitudinal relaxation time $T_1$ for an Si:B ensemble with concentration $n_{\text{B}} \approx 10^{15}~\text{cm}^{-3}$.
A static magnetic field $\overrightarrow{B_0}$ is aligned to the [110] axis of Si crystal for all measurements.


Figures~\ref{Spectra}b and \ref{Spectra}c show spin-echo spectra (see Supplemental material) measured by a Hahn-echo sequence $(\pi/2)_X$--$\tau$--$(\pi)_Y$ (Fig.~\ref{Spectra}b top right) with $(\pi/2)_X$ and $(\pi)_Y$ pulses separated by a time interval $\tau$, where $X, Y$ subscripts indicate the rotation axes in a qubit subsystem.
In the relaxed $^{28}$Si:B sample a narrow spin-echo signal appears at $|\overrightarrow{B_0}|\approx$ 383 mT with a microwave frequency $\omega_{\text{MW}}/2\pi$ of 6.255 GHz (Fig.~\ref{Spectra}b).
We obtain an effective g-factor $|g^{\ast}|$ of $1.17$ by equating $\omega_{\text{MW}}$ with $|g^\ast|\mu_\text{B}|\overrightarrow{B_0}|/\hbar$ where $\mu_\text{B}$ is the Bohr magneton and $\hbar$ is the reduced Planck constant.
This is consistent with $|g^{\ast}|=1.13$ reported in EPR studies of $^{28}$Si:B for $\overrightarrow{B_0}\parallel[110]$ \cite{Neubrand78PSSB,Stegner10PRB}, for the $\{\left|\tfrac{3}{2},+\tfrac{3}{2}\right\rangle,\left|\tfrac{3}{2},+\tfrac{1}{2}\right\rangle\}$ subsystem (black arrow, Fig.~\ref{Concept}a). 
In the strained $^{28}$Si:B sample we observe a spin-echo signal over a broad range of $|g^{\ast}|$ from $2.4$ to $2.6$ at $\omega_{\text{MW}}/2\pi=$ 6.331 GHz (Fig.~\ref{Spectra}c).
No signal is found at $|g^{\ast}|=1.17$ in contrast with the relaxed sample (not shown), ensuring that the sample is properly strained.
We attribute the broad spin-echo signal in the strained sample to the $\{\left|\tfrac{3}{2},+\tfrac{1}{2}\right\rangle,\left|\tfrac{3}{2},-\tfrac{1}{2}\right\rangle\}$ generalised spin (black arrow, Fig.~\ref{Concept}b), since $|g^{\ast}|$ in this range is expected for the configuration of strain and static magnetic field (Fig.~\ref{Spectra}a) \cite{Feher60PRL,Dirksen89JPCM}, as supported by theory (see Supplemental material). 
The broadening of the spin-echo spectrum can be due to the distribution of the $g^{\ast}$ value induced by strain inhomogeneity (see Supplemental material). 
An additional spin-echo peak attributed to dangling bond surface defects (P$_\text{b}$ centres) appears at $|g^{\ast}|\approx2.0$ in both samples.
The relative sharpness of the P$_\text{b}$ signal compared to the strained Si:B signal (Fig.~\ref{Spectra}c) provides evidence that magnetic field inhomogeneity is negligible.

\begin{figure}
[ptb]
\begin{center}
\includegraphics{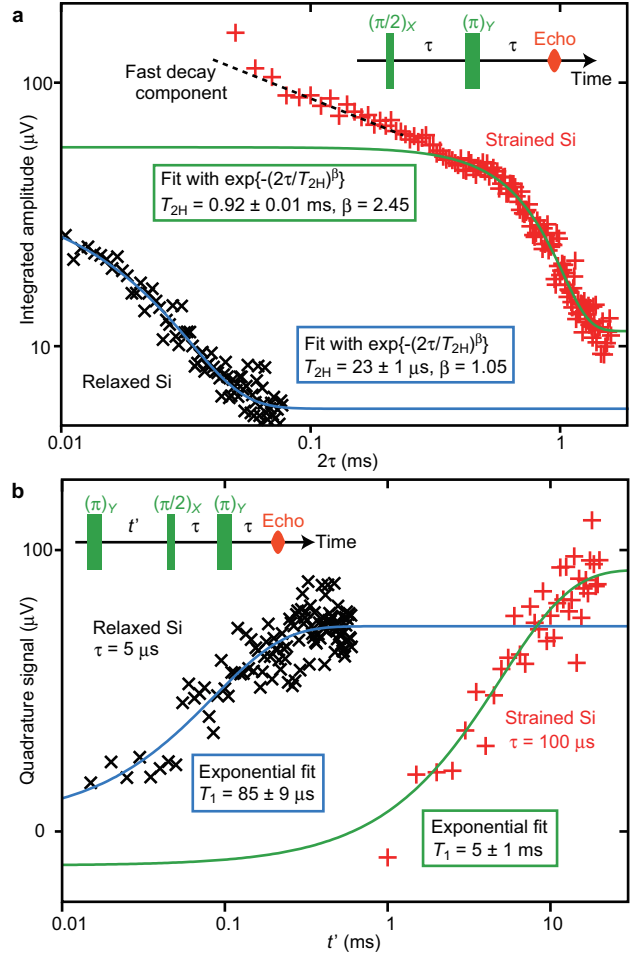}
\caption{Spin-echo signals for the strained (red) and relaxed (black) samples measured by the standard Hahn echo sequence as a function of $\tau$ (\textbf{a}) and measured by the recovery pulse sequence as a function of $t^{\prime}$ (\textbf{b}). The solid curves show fitting functions. The dashed line shows the fast decay component in the strained sample.
}%
\label{T1T2}
\end{center}
\end{figure}

Figure~\ref{T1T2} displays qubit coherence (Fig.~\ref{T1T2}a) and longitudinal relaxation (Fig.~\ref{T1T2}b) measurements in the relaxed and strained samples (black and red symbols, respectively). 
We obtain $T_2$ measured by Hahn-echo decay, $T_{2\text{H}}$, by fitting to a compressed exponential  $A\text{exp}\{-(2\tau/T_{2\text{H}})^\beta\}$, where $\beta$ reveals temporal noise characteristics \cite{Mims68PR,Tyryshkin12NMat}.
For the relaxed sample, fitting results in $T_{2\mathrm{H}}$ of $23\pm1~\mu\text{s}$ and $\beta = 1.05$ (blue curve).
For the strained sample, we find enhanced $T_{2\text{H}}$ of $0.92\pm0.01~\text{ms}$ with $\beta = 2.45$ at $|\overrightarrow{B_0}|=175.7~\text{mT}$ (green curve).
We observe an improvement in $T_{2\text{H}}$ over the full range $g^\ast=2.4$--$2.6$;  $T_{2\text{H}}=0.93\pm0.02,~0.97\pm0.02$, and $0.78\pm0.03~\text{ms}$ and $\beta=3.5,~3.0,$ and $2.9$, for $|\overrightarrow{B_0}|=190.5,~182.0$ and $171.7$ mT, respectively (see Supplemental material).
This improvement compared to the relaxed sample is attributed to the strain-engineered quadrupole coupling as discussed later.
We also find a noticeable component of fast decay ($2\tau\lesssim300~\mu\text{s}$, black dashed line) for the strained sample. 
Because of the random distribution of boron atoms in the sample, the local environments and thus $T_2$ of each generalised spin may differ. 
We posit that the fast decay is due to a generalised-spin subset strongly coupled to decoherence sources such as another closely placed boron atom, impurities, or surface defects, which could be reduced by lower boron concentration, higher crystal purity, or a standard annealing procedure.
The generalised-spin subset well isolated from such decoherence sources is thus responsible for the slow decay with $T_{2\text{H}}$ of $0.92\pm0.01~\text{ms}$.
$T_1$ is measured with an inversion-recovery pulse sequence $(\pi)_Y$--$t^{\prime}$--$(\pi/2)_X$--$\tau$--$(\pi)_Y$ (Fig.~\ref{T1T2}b, top left) consisting of a first inversion pulse, a subsequent Hahn echo pulse sequence and a time interval $t^{\prime}$ between them.
Inversion-recovery signals as a function of $t^{\prime}$ are well fitted by an exponential function $A- B\text{exp}(-t^{\prime}/T_1)$ in both samples (solid curves in Fig~\ref{T1T2}b); we obtain $T_1$ of $85\pm9~\mu\text{s}$ for the relaxed sample and $5\pm1~\text{ms}$ for the strained sample.
The $T_1$ improvement in strained Si:B is consistent with the previous experimental report \cite{Dirksen89JPCM}, attributed to quadrupolar elastic coupling suppressed by strain \cite{Ruskov13PRB}.
We note that $T_{2\mathrm{H}}$ and $T_1$ of the relaxed sample are also longer than those observed in $^{\mathrm{nat}}$Si:B ($\sim2~\mu\text{s}$ and $4~\mu\text{s}$, respectively) \cite{Song11EPL}, which could be explained by a lower temperature than the previous report.


Remarkably, $T_{2\text{H}}$ of the generalised spin $\{\left|\tfrac{3}{2},-\tfrac{1}{2}\right\rangle,\left|\tfrac{3}{2},+\tfrac{1}{2}\right\rangle\}$ in the strained sample is comparable to $T_{2\text{H}}\sim3~\text{ms}$ obtained for $10^{15}$-$\text{cm}^{-3}$ concentration P donors in $^{28}$Si \cite{Tyryshkin12NMat}.
The spin-echo decay curves of the strained sample have $\beta \approx 2.5$--$3.5$ such that decoherence is induced by slow fluctuations (spectral diffusion) \cite{Mims68PR,Tyryshkin12NMat}.
In contrast, an $^{28}$Si:P ensemble has $\beta\approx1$ derived from instantaneous diffusion where spin re-focusing is disrupted by EPR-driven flips of neighbouring spins \cite{Tyryshkin12NMat}.
For strained $^{28}$Si:B, this process has a negligible effect because of g-factor inhomogeneity much larger than the $^{28}$Si:P ensemble.
Without instantaneous diffusion, spectral diffusion in the $^{28}$Si:P ensemble provides $T_{2\text{H}}$ of $\sim100~\mathrm{ms}$.
Importantly, $T_{2\text{H}}$ is also comparable to measured values for state-of-the-art electron-spin qubits defined by $^{28}$Si quantum dots without spin-orbit coupling \cite{Veldhorst14NatureNano}, and exceeds measured values for electron-spin qubits with extrinsic spin-orbit coupling induced by integrated micromagnets \cite{Kawakami14NNano,Yoneda18NNano,Zajac18Science}. 
This demonstrates that the $^{28}$Si:B generalised spin can be as coherent as electron spins in $^{28}$Si.

\begin{figure}
[ptb]
\begin{center}
\includegraphics{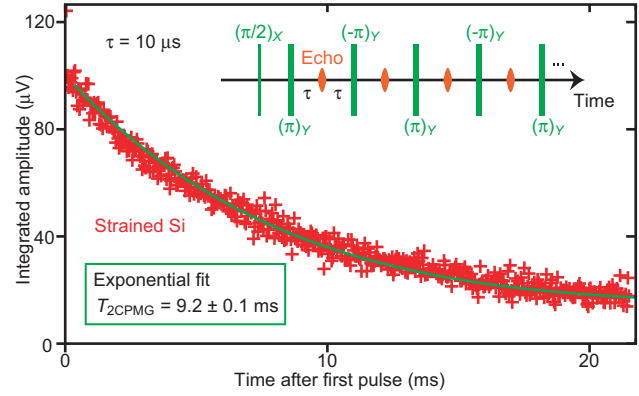}
\caption{Decay of spin echo refocused by the CPMG pulse sequence (top right diagram) in the strained sample as a function of elapsed time after the first $(\pi/2)_X$ pulse, and exponential fit to the data (solid line).
}%
\label{CPMG}
\end{center}
\end{figure}

Decoherence due to spectral diffusion can be ameliorated by dynamical decoupling.
Figure~\ref{CPMG} shows the refocused echo intensity in the strained sample as a function of time after the first $(\pi/2)_X$ pulse, measured by the Carr--Purcell--Meiboom--Gill (CPMG) pulse sequence (top right).
By fitting an exponential function (solid curve) to the spin-echo decay, we obtain $T_{2\text{CPMG}}$ of $9.2\pm0.1~\text{ms}$, that is 10-times longer than $T_{2\text{H}}$ of the strained sample, 400-times longer than $T_{2\text{H}}$ in the relaxed sample and over four orders of magnitude longer than other solid-state systems with intrinsic spin-orbit coupling in previous reports \cite{Nadj-Perge10Nature,Higginbotham14NL,Maurand16NComm}.
Notably, we find that $T_{2\text{CPMG}}$ is very close to the upper bound of $T_2$ set by longitudinal relaxation, $2T_1$, for the strained sample. 
Magnetic field direction and further strain engineering could be used to further improve $T_1$ \cite{Salfi16PRL, AbadilloUriel18APL} and therefore improve $T_{2\text{CPMG}}$ (see Supplemental material).


We now discuss the improvement in hole generalised spin coherence in strained $^{28}$Si:B by analysing electric coupling and the observed Hahn-echo decay curves. 
Generally, qubit dynamics are described by a Hamiltonian $\hat{H}_{\rm qbt}=\tfrac{1}{2}\hbar(\omega_0 + \omega)\hat{\sigma}_Z + \sum_{\alpha=X,Y}\hbar\Omega_\alpha\hat{\sigma}_\alpha$, where $\Omega_\alpha$ is the Rabi frequency and $\hat{\sigma}_\alpha$ ($\alpha=X,Y,Z$) are the Pauli matrices of the qubit subsystem. 
$\omega$ is the change in resonance frequency induced by electric field $\overrightarrow{E}$, expressed as $\omega = \overrightarrow{\chi}\cdot\overrightarrow{E}/\hbar$ with the \textit{longitudinal electric dipole moment} $\overrightarrow{\chi}$ \cite{Beaudoin16Nanotech}.
In any electrically active qubit, decoherence arises from fluctuations in $\omega$, $\delta\omega$, induced by electric field fluctuations $\delta\overrightarrow{E}$, thus mitigated by suppressing $\overrightarrow{\chi}$.
Indeed, $\overrightarrow{\chi}$ for strained Si:B is predicted to be reduced in magnitude by $\hbar\omega_0/2\Delta$ compared to relaxed Si:B for our strain configuration and our magnetic field in the $xy$ plane (see Supplemental material). 
Furthermore, coupling to first order of $\delta\overrightarrow{E}$ only occurs for the $z$-oriented electric fields associated with the bold blue quadrupolar couplings in Fig.~\ref{Concept}c.
This $\overrightarrow{\chi}$ suppression is independently estimated to be $\hbar\omega_0/2\Delta \approx \tfrac{1}{10}$ by comparing the measured $T_1$ in the relaxed and strained samples with perturbation theory (see Supplemental material).
Then, the suppression of $\overrightarrow{\chi}$ by strain reduces $\delta\omega$ attributed to electric fluctuations originating from (i) background electric dipoles by $\hbar\omega_0/2\Delta\sim10^{-1}$ and (ii) flips of neighbouring Si:B hole qubit by $(\hbar\omega_0/2\Delta)^2\sim10^{-2}$ originating from the longitudinal part of the electric dipole-dipole interaction $\sim (1/R^3)|\overrightarrow{\chi}|^2$ for Si:B atoms a distance $R$ apart. 
Another clue in the decoherence processes is a change in the echo-decay exponent from $\beta=1.05$ to $\beta=2.5$--$3.5$, which implies that the characteristic time scale of dominant fluctuations $\tau_c$ changes from $\tau_c > T_{2\text{H}}$, to $\tau_c \lesssim T_{2\text{H}}$ \cite{Mims68PR}. 
The $T_{2\text{H}}$ improvement arises then from suppression in longitudinal coupling to fast electric fluctuators, such that slower fluctuators emerge as the dominant decoherence mechanism. 
One prominent possibility is that the dominant fluctuators in the strained sample become magnetic in origin under the reduced longitudinal electric coupling. 
An increase in $\tau_c$ for the dominant electric fluctuators in the strained sample cannot be ruled out; given that the dominant electric fluctuator is the Si:B hole qubit subsystem not excited by EPR pulses, their spin-flip processes induced by longitudinal relaxation and spin-pair flip-flop are both slowed down by reduced ${T_1}^{-1}$ and transverse dipole-dipole coupling, respectively. 
In either case, the reduced longitudinal electric dipole is a key ingredient to improve coherence.


Here we discuss two key properties, electrical controllability and the intrinsic EPR linewidth, of the generalised spin qubit based on a theoretical model developed in Supplemental material.
Electrical controllability of a qubit is characterised by the \textit{transverse electric dipole moment} $\overrightarrow{v_{\alpha}}$ \cite{Beaudoin16Nanotech}, which provides Rabi frequency $\Omega_\alpha=\overrightarrow{v_{\alpha}}\cdot\overrightarrow{E}/\hbar$.
In the lowest order of electric field, $|\overrightarrow{v_{\alpha}}|$ is reported to be 0.26 Debye for the relaxed Si:B system \cite{Kopf92PRL}, suppressed by $\sqrt{3}\hbar\omega_0/2\Delta$ in our strained Si:B system.
Consequently, $\Omega_\alpha/2\pi$ of $10~\text{MHz}$ will be available in the strained system with realistic oscillating electric fields of $40~\text{kV}/\text{m}$.
The intrinsic linewidth is determined by electric field noise and the longitudinal dipole coupling $\overrightarrow{\chi}$.
The magnitude of $\overrightarrow{\chi}$ is suppressed by a factor of $\hbar\omega_0/2\Delta$ for strained Si:B compared to relaxed Si:B (0.26 Debye \cite{Kopf92PRL}) in a magnetic field along the [110] direction. 
An amplitude of electric field noise has been reported $\sim10~\mathrm{V/m}$ in interface-defined silicon quantum dots \cite{Yoneda18NNano}, thus implying the intrinsic linewidth of $\sim2~\mathrm{kHz}$ in strained Si:B comparable to EPR linewidths of state-of-the-art $^{28}$Si electron spin qubits without external sources of spin-orbit coupling \cite{Muhonen14NNano,Veldhorst14NatureNano}.
We also note that smaller $|\overrightarrow{\chi}|$ and thus longer $T_2$ can be available by aligning the magnetic field to the [100] direction without influencing $\overrightarrow{v_{\alpha}}$.
These observations imply that acceptor-bound holes embedded into silicon field-effect transistor devices \cite{vanderHeijden17arXiv} could offer long $T_2$ and spin-orbit functionality usable for quantum manipulations.


Together with long $T_2$, several mechanisms could be employed to realise hybrid spin-photon systems or long-range spin-spin interactions via photons in superconducting microwave resonators. 
The key strategy will be to engineer the spin-photon interaction to realise the strong coupling regime without increasing decoherence. 
This could be achieved by periodically modulating the transverse or longitudinal couplings to enhance the spin-photon interaction \cite{Lambert18PRB}. 
Indeed, an oscillating control electric field $\overrightarrow{E}_c(t)$ can be used to periodically modulate $\overrightarrow{v_\alpha}(\overrightarrow{E}_c(t))$ and $\overrightarrow{\chi}(\overrightarrow{E}_c(t)$) via the second-order effect of electric field attributed to quadrupolar coupling of hole generalised spins (see Supplemental material).
Alternatively, the transverse coupling to $x$- and $y$-oriented electric fields could be statically enhanced using a $z$-oriented electric field and interface via a Rashba-like quadrupolar interaction at a sweet spot with long $T_2$ \cite{Salfi16PRL}. 
We note that $\overrightarrow{E}$ along the $z$ axis enhances $\overrightarrow{v_\alpha}(\overrightarrow{E})$ without changing the $\overrightarrow{\chi}(\overrightarrow{E})$ up to higher order terms for generalised spins (see Supplemental material).
This means that the generalised-spin qubit allows to control the transverse coupling without increasing the longitudinal coupling by $\overrightarrow{E}$ along $z$.
Phonon coupling stands out as an alternative qubit coupling mechanism \cite{Ruskov13PRB}, and the coherence properties we have verified make Si:B hole generalised spins interesting candidates for phonon coupled hybrid systems using silicon mechanical resonators.  
Systems where holes are allowed to tunnel between neighbouring quantum dot or Si:B sites \cite{vanderHeijden17arXiv} will experience spin-orbit coupling, which could be useful to realise exotic spin-orbit coupled states \cite{Kato04Science,Sau10PRL}. 
It should be possible to combine these spin-orbit functionalities with long $T_2$ because the transverse coupling could be engineered without increasing decoherence induced by the longitudinal coupling to electric fields.


We have experimentally established long $T_2$ in generalised hole spins bound to acceptor atoms in mechanically strained Si where total angular momentum $J$, and not real spin $S$, is a good quantum number.
Our measured $T_2$ times are similar to state-of-the-art results for systems such as electrons bound to Si:P donors and Si quantum dots, and three to four orders of magnitude longer than other solid-state systems with spin-orbit coupling. 
These observations open up a new and promising pathway to use spin-orbit coupling available in holes to engineer new kinds of highly coherent hybrid quantum systems and to achieve long distance spin qubit coupling for single atom qubits in silicon.


We acknowledge that this work was supported by the ARC Centre of Excellence for Quantum Computation and Communication Technology (CE170100012), in part by the U.S. Army Research Office (W911NF-08-1-0527). T.K. acknowledges support from the Tohoku University Graduate Program in Spintronics. J.S. acknowledges support from an ARC DECRA fellowship (DE160101490). M.Y.S. acknowledges a Laureate Fellowship.  The authors thank Mike Thewalt for the $^{28}$Si sample.


\clearpage


\clearpage


\clearpage


\clearpage

\clearpage
\bibliography{Manuscript_arXiv}

\begin{thebibliography}{35}%
\makeatletter
\providecommand \@ifxundefined [1]{%
 \@ifx{#1\undefined}
}%
\providecommand \@ifnum [1]{%
 \ifnum #1\expandafter \@firstoftwo
 \else \expandafter \@secondoftwo
 \fi
}%
\providecommand \@ifx [1]{%
 \ifx #1\expandafter \@firstoftwo
 \else \expandafter \@secondoftwo
 \fi
}%
\providecommand \natexlab [1]{#1}%
\providecommand \enquote  [1]{``#1''}%
\providecommand \bibnamefont  [1]{#1}%
\providecommand \bibfnamefont [1]{#1}%
\providecommand \citenamefont [1]{#1}%
\providecommand \href@noop [0]{\@secondoftwo}%
\providecommand \href [0]{\begingroup \@sanitize@url \@href}%
\providecommand \@href[1]{\@@startlink{#1}\@@href}%
\providecommand \@@href[1]{\endgroup#1\@@endlink}%
\providecommand \@sanitize@url [0]{\catcode `\\12\catcode `\$12\catcode
  `\&12\catcode `\#12\catcode `\^12\catcode `\_12\catcode `\%12\relax}%
\providecommand \@@startlink[1]{}%
\providecommand \@@endlink[0]{}%
\providecommand \url  [0]{\begingroup\@sanitize@url \@url }%
\providecommand \@url [1]{\endgroup\@href {#1}{\urlprefix }}%
\providecommand \urlprefix  [0]{URL }%
\providecommand \Eprint [0]{\href }%
\providecommand \doibase [0]{http://dx.doi.org/}%
\providecommand \selectlanguage [0]{\@gobble}%
\providecommand \bibinfo  [0]{\@secondoftwo}%
\providecommand \bibfield  [0]{\@secondoftwo}%
\providecommand \translation [1]{[#1]}%
\providecommand \BibitemOpen [0]{}%
\providecommand \bibitemStop [0]{}%
\providecommand \bibitemNoStop [0]{.\EOS\space}%
\providecommand \EOS [0]{\spacefactor3000\relax}%
\providecommand \BibitemShut  [1]{\csname bibitem#1\endcsname}%
\let\auto@bib@innerbib\@empty
\bibitem [{\citenamefont {Ladd}\ \emph {et~al.}(2010)\citenamefont {Ladd},
  \citenamefont {Jelezko}, \citenamefont {Laflamme}, \citenamefont {Nakamura},
  \citenamefont {Monroe},\ and\ \citenamefont {O'Brien}}]{Ladd10Nature}%
  \BibitemOpen
  \bibfield  {author} {\bibinfo {author} {\bibfnamefont {T.~D.}\ \bibnamefont
  {Ladd}}, \bibinfo {author} {\bibfnamefont {F.}~\bibnamefont {Jelezko}},
  \bibinfo {author} {\bibfnamefont {R.}~\bibnamefont {Laflamme}}, \bibinfo
  {author} {\bibfnamefont {Y.}~\bibnamefont {Nakamura}}, \bibinfo {author}
  {\bibfnamefont {C.}~\bibnamefont {Monroe}}, \ and\ \bibinfo {author}
  {\bibfnamefont {J.~L.}\ \bibnamefont {O'Brien}},\ }\href@noop {} {\bibfield
  {journal} {\bibinfo  {journal} {Nature}\ }\textbf {\bibinfo {volume} {464}},\
  \bibinfo {pages} {45} (\bibinfo {year} {2010})}\BibitemShut {NoStop}%
\bibitem [{\citenamefont {Xiang}\ \emph {et~al.}(2013)\citenamefont {Xiang},
  \citenamefont {Ashhab}, \citenamefont {You},\ and\ \citenamefont
  {Nori}}]{Xiang13RMP}%
  \BibitemOpen
  \bibfield  {author} {\bibinfo {author} {\bibfnamefont {Z.-L.}\ \bibnamefont
  {Xiang}}, \bibinfo {author} {\bibfnamefont {S.}~\bibnamefont {Ashhab}},
  \bibinfo {author} {\bibfnamefont {J.~Q.}\ \bibnamefont {You}}, \ and\
  \bibinfo {author} {\bibfnamefont {F.}~\bibnamefont {Nori}},\ }\href {\doibase
  10.1103/RevModPhys.85.623} {\bibfield  {journal} {\bibinfo  {journal} {Rev.
  Mod. Phys.}\ }\textbf {\bibinfo {volume} {85}},\ \bibinfo {pages} {623}
  (\bibinfo {year} {2013})}\BibitemShut {NoStop}%
\bibitem [{\citenamefont {Georgescu}\ \emph {et~al.}(2014)\citenamefont
  {Georgescu}, \citenamefont {Ashhab},\ and\ \citenamefont
  {Nori}}]{Georgescu14RMP}%
  \BibitemOpen
  \bibfield  {author} {\bibinfo {author} {\bibfnamefont {I.~M.}\ \bibnamefont
  {Georgescu}}, \bibinfo {author} {\bibfnamefont {S.}~\bibnamefont {Ashhab}}, \
  and\ \bibinfo {author} {\bibfnamefont {F.}~\bibnamefont {Nori}},\ }\href
  {\doibase 10.1103/RevModPhys.86.153} {\bibfield  {journal} {\bibinfo
  {journal} {Rev. Mod. Phys.}\ }\textbf {\bibinfo {volume} {86}},\ \bibinfo
  {pages} {153} (\bibinfo {year} {2014})}\BibitemShut {NoStop}%
\bibitem [{\citenamefont {Kato}\ \emph {et~al.}(2004)\citenamefont {Kato},
  \citenamefont {Myers}, \citenamefont {Gossard},\ and\ \citenamefont
  {Awschalom}}]{Kato04Science}%
  \BibitemOpen
  \bibfield  {author} {\bibinfo {author} {\bibfnamefont {Y.~K.}\ \bibnamefont
  {Kato}}, \bibinfo {author} {\bibfnamefont {R.~C.}\ \bibnamefont {Myers}},
  \bibinfo {author} {\bibfnamefont {A.~C.}\ \bibnamefont {Gossard}}, \ and\
  \bibinfo {author} {\bibfnamefont {D.~D.}\ \bibnamefont {Awschalom}},\
  }\href@noop {} {\bibfield  {journal} {\bibinfo  {journal} {Science}\ }\textbf
  {\bibinfo {volume} {306}},\ \bibinfo {pages} {1910} (\bibinfo {year}
  {2004})}\BibitemShut {NoStop}%
\bibitem [{\citenamefont {Sau}\ \emph {et~al.}(2010)\citenamefont {Sau},
  \citenamefont {Lutchyn}, \citenamefont {Tewari},\ and\ \citenamefont
  {Das~Sarma}}]{Sau10PRL}%
  \BibitemOpen
  \bibfield  {author} {\bibinfo {author} {\bibfnamefont {J.~D.}\ \bibnamefont
  {Sau}}, \bibinfo {author} {\bibfnamefont {R.~M.}\ \bibnamefont {Lutchyn}},
  \bibinfo {author} {\bibfnamefont {S.}~\bibnamefont {Tewari}}, \ and\ \bibinfo
  {author} {\bibfnamefont {S.}~\bibnamefont {Das~Sarma}},\ }\href@noop {}
  {\bibfield  {journal} {\bibinfo  {journal} {Phys. Rev. Lett.}\ }\textbf
  {\bibinfo {volume} {104}},\ \bibinfo {pages} {040502} (\bibinfo {year}
  {2010})}\BibitemShut {NoStop}%
\bibitem [{\citenamefont {Nadj-Perge}\ \emph {et~al.}(2010)\citenamefont
  {Nadj-Perge}, \citenamefont {Frolov}, \citenamefont {Bakkers},\ and\
  \citenamefont {Kouwenhoven}}]{Nadj-Perge10Nature}%
  \BibitemOpen
  \bibfield  {author} {\bibinfo {author} {\bibfnamefont {S.}~\bibnamefont
  {Nadj-Perge}}, \bibinfo {author} {\bibfnamefont {S.~M.}\ \bibnamefont
  {Frolov}}, \bibinfo {author} {\bibfnamefont {E.~P. A.~M.}\ \bibnamefont
  {Bakkers}}, \ and\ \bibinfo {author} {\bibfnamefont {L.~P.}\ \bibnamefont
  {Kouwenhoven}},\ }\href@noop {} {\bibfield  {journal} {\bibinfo  {journal}
  {Nature}\ }\textbf {\bibinfo {volume} {468}},\ \bibinfo {pages} {1084}
  (\bibinfo {year} {2010})}\BibitemShut {NoStop}%
\bibitem [{\citenamefont {Song}\ and\ \citenamefont
  {Golding}(2011)}]{Song11EPL}%
  \BibitemOpen
  \bibfield  {author} {\bibinfo {author} {\bibfnamefont {Y.~P.}\ \bibnamefont
  {Song}}\ and\ \bibinfo {author} {\bibfnamefont {B.}~\bibnamefont {Golding}},\
  }\href@noop {} {\bibfield  {journal} {\bibinfo  {journal} {Europhys. Lett.}\
  }\textbf {\bibinfo {volume} {95}},\ \bibinfo {pages} {47004} (\bibinfo {year}
  {2011})}\BibitemShut {NoStop}%
\bibitem [{\citenamefont {Higginbotham}\ \emph {et~al.}(2014)\citenamefont
  {Higginbotham}, \citenamefont {Larsen}, \citenamefont {Yao}, \citenamefont
  {Yan}, \citenamefont {Lieber}, \citenamefont {Marcus},\ and\ \citenamefont
  {Kuemmeth}}]{Higginbotham14NL}%
  \BibitemOpen
  \bibfield  {author} {\bibinfo {author} {\bibfnamefont {A.~P.}\ \bibnamefont
  {Higginbotham}}, \bibinfo {author} {\bibfnamefont {T.~W.}\ \bibnamefont
  {Larsen}}, \bibinfo {author} {\bibfnamefont {J.}~\bibnamefont {Yao}},
  \bibinfo {author} {\bibfnamefont {H.}~\bibnamefont {Yan}}, \bibinfo {author}
  {\bibfnamefont {C.~M.}\ \bibnamefont {Lieber}}, \bibinfo {author}
  {\bibfnamefont {C.~M.}\ \bibnamefont {Marcus}}, \ and\ \bibinfo {author}
  {\bibfnamefont {F.}~\bibnamefont {Kuemmeth}},\ }\href@noop {} {\bibfield
  {journal} {\bibinfo  {journal} {Nano Lett.}\ }\textbf {\bibinfo {volume}
  {14}},\ \bibinfo {pages} {3582} (\bibinfo {year} {2014})}\BibitemShut
  {NoStop}%
\bibitem [{\citenamefont {Maurand}\ \emph {et~al.}(2016)\citenamefont
  {Maurand}, \citenamefont {Jehl}, \citenamefont {Kotekar-Patil}, \citenamefont
  {Corna}, \citenamefont {Bohuslavskyi}, \citenamefont {LaviChan~ville},
  \citenamefont {Hutin}, \citenamefont {Barraud}, \citenamefont {Vinet},
  \citenamefont {Sanquer},\ and\ \citenamefont
  {De~Franceschi}}]{Maurand16NComm}%
  \BibitemOpen
  \bibfield  {author} {\bibinfo {author} {\bibfnamefont {R.}~\bibnamefont
  {Maurand}}, \bibinfo {author} {\bibfnamefont {X.}~\bibnamefont {Jehl}},
  \bibinfo {author} {\bibfnamefont {D.}~\bibnamefont {Kotekar-Patil}}, \bibinfo
  {author} {\bibfnamefont {A.}~\bibnamefont {Corna}}, \bibinfo {author}
  {\bibfnamefont {H.}~\bibnamefont {Bohuslavskyi}}, \bibinfo {author}
  {\bibfnamefont {R.}~\bibnamefont {LaviChan~ville}}, \bibinfo {author}
  {\bibfnamefont {L.}~\bibnamefont {Hutin}}, \bibinfo {author} {\bibfnamefont
  {S.}~\bibnamefont {Barraud}}, \bibinfo {author} {\bibfnamefont
  {M.}~\bibnamefont {Vinet}}, \bibinfo {author} {\bibfnamefont
  {M.}~\bibnamefont {Sanquer}}, \ and\ \bibinfo {author} {\bibfnamefont
  {S.}~\bibnamefont {De~Franceschi}},\ }\href@noop {} {\bibfield  {journal}
  {\bibinfo  {journal} {Nature Commun.}\ }\textbf {\bibinfo {volume} {7}},\
  \bibinfo {pages} {13575} (\bibinfo {year} {2016})},\ \bibinfo {note}
  {article}\BibitemShut {NoStop}%
\bibitem [{\citenamefont {Tyryshkin}\ \emph {et~al.}(2012)\citenamefont
  {Tyryshkin}, \citenamefont {Tojo}, \citenamefont {Morton}, \citenamefont
  {Riemann}, \citenamefont {Abrosimov}, \citenamefont {Becker}, \citenamefont
  {Pohl}, \citenamefont {Schenkel}, \citenamefont {Thewalt}, \citenamefont
  {Itoh},\ and\ \citenamefont {Lyon}}]{Tyryshkin12NMat}%
  \BibitemOpen
  \bibfield  {author} {\bibinfo {author} {\bibfnamefont {A.~M.}\ \bibnamefont
  {Tyryshkin}}, \bibinfo {author} {\bibfnamefont {S.}~\bibnamefont {Tojo}},
  \bibinfo {author} {\bibfnamefont {J.~J.~L.}\ \bibnamefont {Morton}}, \bibinfo
  {author} {\bibfnamefont {H.}~\bibnamefont {Riemann}}, \bibinfo {author}
  {\bibfnamefont {N.~V.}\ \bibnamefont {Abrosimov}}, \bibinfo {author}
  {\bibfnamefont {P.}~\bibnamefont {Becker}}, \bibinfo {author} {\bibfnamefont
  {H.-J.}\ \bibnamefont {Pohl}}, \bibinfo {author} {\bibfnamefont
  {T.}~\bibnamefont {Schenkel}}, \bibinfo {author} {\bibfnamefont {M.~L.~W.}\
  \bibnamefont {Thewalt}}, \bibinfo {author} {\bibfnamefont {K.~M.}\
  \bibnamefont {Itoh}}, \ and\ \bibinfo {author} {\bibfnamefont {S.~A.}\
  \bibnamefont {Lyon}},\ }\href@noop {} {\bibfield  {journal} {\bibinfo
  {journal} {Nature Mater.}\ }\textbf {\bibinfo {volume} {11}},\ \bibinfo
  {pages} {143} (\bibinfo {year} {2012})}\BibitemShut {NoStop}%
\bibitem [{\citenamefont {Muhonen}\ \emph {et~al.}(2014)\citenamefont
  {Muhonen}, \citenamefont {Dehollain}, \citenamefont {Laucht}, \citenamefont
  {Hudson}, \citenamefont {Kalra}, \citenamefont {Sekiguchi}, \citenamefont
  {Itoh}, \citenamefont {Jamieson}, \citenamefont {McCallum}, \citenamefont
  {Dzurak},\ and\ \citenamefont {Morello}}]{Muhonen14NNano}%
  \BibitemOpen
  \bibfield  {author} {\bibinfo {author} {\bibfnamefont {J.~T.}\ \bibnamefont
  {Muhonen}}, \bibinfo {author} {\bibfnamefont {J.~P.}\ \bibnamefont
  {Dehollain}}, \bibinfo {author} {\bibfnamefont {A.}~\bibnamefont {Laucht}},
  \bibinfo {author} {\bibfnamefont {F.~E.}\ \bibnamefont {Hudson}}, \bibinfo
  {author} {\bibfnamefont {R.}~\bibnamefont {Kalra}}, \bibinfo {author}
  {\bibfnamefont {T.}~\bibnamefont {Sekiguchi}}, \bibinfo {author}
  {\bibfnamefont {K.~M.}\ \bibnamefont {Itoh}}, \bibinfo {author}
  {\bibfnamefont {D.~N.}\ \bibnamefont {Jamieson}}, \bibinfo {author}
  {\bibfnamefont {J.~C.}\ \bibnamefont {McCallum}}, \bibinfo {author}
  {\bibfnamefont {A.~S.}\ \bibnamefont {Dzurak}}, \ and\ \bibinfo {author}
  {\bibfnamefont {A.}~\bibnamefont {Morello}},\ }\href@noop {} {\bibfield
  {journal} {\bibinfo  {journal} {Nature Nanotechnol.}\ }\textbf {\bibinfo
  {volume} {9}},\ \bibinfo {pages} {986} (\bibinfo {year} {2014})}\BibitemShut
  {NoStop}%
\bibitem [{\citenamefont {Veldhorst}\ \emph {et~al.}(2014)\citenamefont
  {Veldhorst}, \citenamefont {Hwang}, \citenamefont {Yang}, \citenamefont
  {Leenstra}, \citenamefont {de~Ronde}, \citenamefont {Dehollain},
  \citenamefont {Muhonen}, \citenamefont {Hudson}, \citenamefont {Itoh},
  \citenamefont {Morello},\ and\ \citenamefont
  {Dzurak}}]{Veldhorst14NatureNano}%
  \BibitemOpen
  \bibfield  {author} {\bibinfo {author} {\bibfnamefont {M.}~\bibnamefont
  {Veldhorst}}, \bibinfo {author} {\bibfnamefont {J.~C.~C.}\ \bibnamefont
  {Hwang}}, \bibinfo {author} {\bibfnamefont {C.~H.}\ \bibnamefont {Yang}},
  \bibinfo {author} {\bibfnamefont {A.~W.}\ \bibnamefont {Leenstra}}, \bibinfo
  {author} {\bibfnamefont {B.}~\bibnamefont {de~Ronde}}, \bibinfo {author}
  {\bibfnamefont {J.~P.}\ \bibnamefont {Dehollain}}, \bibinfo {author}
  {\bibfnamefont {J.~T.}\ \bibnamefont {Muhonen}}, \bibinfo {author}
  {\bibfnamefont {F.~E.}\ \bibnamefont {Hudson}}, \bibinfo {author}
  {\bibfnamefont {K.~M.}\ \bibnamefont {Itoh}}, \bibinfo {author}
  {\bibfnamefont {A.}~\bibnamefont {Morello}}, \ and\ \bibinfo {author}
  {\bibfnamefont {A.~S.}\ \bibnamefont {Dzurak}},\ }\href@noop {} {\bibfield
  {journal} {\bibinfo  {journal} {Nature Nanotech}\ }\textbf {\bibinfo {volume}
  {9}},\ \bibinfo {pages} {981} (\bibinfo {year} {2014})}\BibitemShut {NoStop}%
\bibitem [{\citenamefont {Beaudoin}\ \emph {et~al.}(2016)\citenamefont
  {Beaudoin}, \citenamefont {Lachance-Quirion}, \citenamefont {Coish},\ and\
  \citenamefont {Pioro-Ladri\`{e}re}}]{Beaudoin16Nanotech}%
  \BibitemOpen
  \bibfield  {author} {\bibinfo {author} {\bibfnamefont {F.}~\bibnamefont
  {Beaudoin}}, \bibinfo {author} {\bibfnamefont {D.}~\bibnamefont
  {Lachance-Quirion}}, \bibinfo {author} {\bibfnamefont {W.~A.}\ \bibnamefont
  {Coish}}, \ and\ \bibinfo {author} {\bibfnamefont {M.}~\bibnamefont
  {Pioro-Ladri\`{e}re}},\ }\href@noop {} {\bibfield  {journal} {\bibinfo
  {journal} {Nanotech.}\ }\textbf {\bibinfo {volume} {27}},\ \bibinfo {pages}
  {464003} (\bibinfo {year} {2016})}\BibitemShut {NoStop}%
\bibitem [{\citenamefont {Luttinger}\ and\ \citenamefont
  {Kohn}(1955)}]{Luttinger55PR}%
  \BibitemOpen
  \bibfield  {author} {\bibinfo {author} {\bibfnamefont {J.~M.}\ \bibnamefont
  {Luttinger}}\ and\ \bibinfo {author} {\bibfnamefont {W.}~\bibnamefont
  {Kohn}},\ }\href@noop {} {\bibfield  {journal} {\bibinfo  {journal} {Phys.
  Rev.}\ }\textbf {\bibinfo {volume} {97}},\ \bibinfo {pages} {869} (\bibinfo
  {year} {1955})}\BibitemShut {NoStop}%
\bibitem [{\citenamefont {Winkler}(2004)}]{Winkler04PRB}%
  \BibitemOpen
  \bibfield  {author} {\bibinfo {author} {\bibfnamefont {R.}~\bibnamefont
  {Winkler}},\ }\href@noop {} {\bibfield  {journal} {\bibinfo  {journal} {Phys.
  Rev. B}\ }\textbf {\bibinfo {volume} {70}},\ \bibinfo {pages} {125301}
  (\bibinfo {year} {2004})}\BibitemShut {NoStop}%
\bibitem [{\citenamefont {Kim}\ \emph {et~al.}(2014)\citenamefont {Kim},
  \citenamefont {Shi}, \citenamefont {Simmons}, \citenamefont {Ward},
  \citenamefont {Prance}, \citenamefont {Koh}, \citenamefont {Gamble},
  \citenamefont {Savage}, \citenamefont {Lagally}, \citenamefont {Friesen},
  \citenamefont {Coppersmith},\ and\ \citenamefont {Eriksson}}]{Kim14Nature}%
  \BibitemOpen
  \bibfield  {author} {\bibinfo {author} {\bibfnamefont {D.}~\bibnamefont
  {Kim}}, \bibinfo {author} {\bibfnamefont {Z.}~\bibnamefont {Shi}}, \bibinfo
  {author} {\bibfnamefont {C.~B.}\ \bibnamefont {Simmons}}, \bibinfo {author}
  {\bibfnamefont {D.~R.}\ \bibnamefont {Ward}}, \bibinfo {author}
  {\bibfnamefont {J.~R.}\ \bibnamefont {Prance}}, \bibinfo {author}
  {\bibfnamefont {T.~S.}\ \bibnamefont {Koh}}, \bibinfo {author} {\bibfnamefont
  {J.~K.}\ \bibnamefont {Gamble}}, \bibinfo {author} {\bibfnamefont {D.~E.}\
  \bibnamefont {Savage}}, \bibinfo {author} {\bibfnamefont {M.~G.}\
  \bibnamefont {Lagally}}, \bibinfo {author} {\bibfnamefont {M.}~\bibnamefont
  {Friesen}}, \bibinfo {author} {\bibfnamefont {S.~N.}\ \bibnamefont
  {Coppersmith}}, \ and\ \bibinfo {author} {\bibfnamefont {M.~A.}\ \bibnamefont
  {Eriksson}},\ }\href@noop {} {\bibfield  {journal} {\bibinfo  {journal}
  {Nature}\ }\textbf {\bibinfo {volume} {511}},\ \bibinfo {pages} {70}
  (\bibinfo {year} {2014})}\BibitemShut {NoStop}%
\bibitem [{\citenamefont {Kawakami}\ \emph {et~al.}(2014)\citenamefont
  {Kawakami}, \citenamefont {Scarlino}, \citenamefont {Ward}, \citenamefont
  {Braakman}, \citenamefont {Savage}, \citenamefont {Lagally}, \citenamefont
  {Friesen}, \citenamefont {Coppersmith}, \citenamefont {Eriksson},\ and\
  \citenamefont {Vandersypen}}]{Kawakami14NNano}%
  \BibitemOpen
  \bibfield  {author} {\bibinfo {author} {\bibfnamefont {E.}~\bibnamefont
  {Kawakami}}, \bibinfo {author} {\bibfnamefont {P.}~\bibnamefont {Scarlino}},
  \bibinfo {author} {\bibfnamefont {D.~R.}\ \bibnamefont {Ward}}, \bibinfo
  {author} {\bibfnamefont {F.~R.}\ \bibnamefont {Braakman}}, \bibinfo {author}
  {\bibfnamefont {D.~E.}\ \bibnamefont {Savage}}, \bibinfo {author}
  {\bibfnamefont {M.~G.}\ \bibnamefont {Lagally}}, \bibinfo {author}
  {\bibfnamefont {M.}~\bibnamefont {Friesen}}, \bibinfo {author} {\bibfnamefont
  {S.~N.}\ \bibnamefont {Coppersmith}}, \bibinfo {author} {\bibfnamefont
  {M.~A.}\ \bibnamefont {Eriksson}}, \ and\ \bibinfo {author} {\bibfnamefont
  {L.~M.~K.}\ \bibnamefont {Vandersypen}},\ }\href@noop {} {\bibfield
  {journal} {\bibinfo  {journal} {Nature Nanotechnol.}\ }\textbf {\bibinfo
  {volume} {9}},\ \bibinfo {pages} {666} (\bibinfo {year} {2014})}\BibitemShut
  {NoStop}%
\bibitem [{\citenamefont {Yoneda}\ \emph {et~al.}(2018)\citenamefont {Yoneda},
  \citenamefont {Takeda}, \citenamefont {Otsuka}, \citenamefont {Nakajima},
  \citenamefont {Delbecq}, \citenamefont {Allison}, \citenamefont {Honda},
  \citenamefont {Kodera}, \citenamefont {Oda}, \citenamefont {Hoshi},
  \citenamefont {Usami}, \citenamefont {Itoh},\ and\ \citenamefont
  {Tarucha}}]{Yoneda18NNano}%
  \BibitemOpen
  \bibfield  {author} {\bibinfo {author} {\bibfnamefont {J.}~\bibnamefont
  {Yoneda}}, \bibinfo {author} {\bibfnamefont {K.}~\bibnamefont {Takeda}},
  \bibinfo {author} {\bibfnamefont {T.}~\bibnamefont {Otsuka}}, \bibinfo
  {author} {\bibfnamefont {T.}~\bibnamefont {Nakajima}}, \bibinfo {author}
  {\bibfnamefont {M.~R.}\ \bibnamefont {Delbecq}}, \bibinfo {author}
  {\bibfnamefont {G.}~\bibnamefont {Allison}}, \bibinfo {author} {\bibfnamefont
  {T.}~\bibnamefont {Honda}}, \bibinfo {author} {\bibfnamefont
  {T.}~\bibnamefont {Kodera}}, \bibinfo {author} {\bibfnamefont
  {S.}~\bibnamefont {Oda}}, \bibinfo {author} {\bibfnamefont {Y.}~\bibnamefont
  {Hoshi}}, \bibinfo {author} {\bibfnamefont {N.}~\bibnamefont {Usami}},
  \bibinfo {author} {\bibfnamefont {K.~M.}\ \bibnamefont {Itoh}}, \ and\
  \bibinfo {author} {\bibfnamefont {S.}~\bibnamefont {Tarucha}},\ }\href@noop
  {} {\bibfield  {journal} {\bibinfo  {journal} {Nature Nanotechnol.}\ }\textbf
  {\bibinfo {volume} {13}},\ \bibinfo {pages} {102} (\bibinfo {year}
  {2018})}\BibitemShut {NoStop}%
\bibitem [{\citenamefont {Zajac}\ \emph {et~al.}(2018)\citenamefont {Zajac},
  \citenamefont {Sigillito}, \citenamefont {Russ}, \citenamefont {Borjans},
  \citenamefont {Taylor}, \citenamefont {Burkard},\ and\ \citenamefont
  {Petta}}]{Zajac18Science}%
  \BibitemOpen
  \bibfield  {author} {\bibinfo {author} {\bibfnamefont {D.~M.}\ \bibnamefont
  {Zajac}}, \bibinfo {author} {\bibfnamefont {A.~J.}\ \bibnamefont
  {Sigillito}}, \bibinfo {author} {\bibfnamefont {M.}~\bibnamefont {Russ}},
  \bibinfo {author} {\bibfnamefont {F.}~\bibnamefont {Borjans}}, \bibinfo
  {author} {\bibfnamefont {J.~M.}\ \bibnamefont {Taylor}}, \bibinfo {author}
  {\bibfnamefont {G.}~\bibnamefont {Burkard}}, \ and\ \bibinfo {author}
  {\bibfnamefont {J.~R.}\ \bibnamefont {Petta}},\ }\href@noop {} {\bibfield
  {journal} {\bibinfo  {journal} {Science}\ }\textbf {\bibinfo {volume}
  {359}},\ \bibinfo {pages} {439} (\bibinfo {year} {2018})}\BibitemShut
  {NoStop}%
\bibitem [{\citenamefont {Salfi}\ \emph {et~al.}(2016)\citenamefont {Salfi},
  \citenamefont {Mol}, \citenamefont {Culcer},\ and\ \citenamefont
  {Rogge}}]{Salfi16PRL}%
  \BibitemOpen
  \bibfield  {author} {\bibinfo {author} {\bibfnamefont {J.}~\bibnamefont
  {Salfi}}, \bibinfo {author} {\bibfnamefont {J.~A.}\ \bibnamefont {Mol}},
  \bibinfo {author} {\bibfnamefont {D.}~\bibnamefont {Culcer}}, \ and\ \bibinfo
  {author} {\bibfnamefont {S.}~\bibnamefont {Rogge}},\ }\href@noop {}
  {\bibfield  {journal} {\bibinfo  {journal} {Phys. Rev. Lett.}\ }\textbf
  {\bibinfo {volume} {116}},\ \bibinfo {pages} {246801} (\bibinfo {year}
  {2016})}\BibitemShut {NoStop}%
\bibitem [{\citenamefont {van~der Heijden}\ \emph {et~al.}(2017)\citenamefont
  {van~der Heijden}, \citenamefont {Kobayashi}, \citenamefont {House},
  \citenamefont {Slafi}, \citenamefont {Barraud}, \citenamefont {Lavieville},
  \citenamefont {Simmons},\ and\ \citenamefont {Rogge}}]{vanderHeijden17arXiv}%
  \BibitemOpen
  \bibfield  {author} {\bibinfo {author} {\bibfnamefont {J.}~\bibnamefont
  {van~der Heijden}}, \bibinfo {author} {\bibfnamefont {T.}~\bibnamefont
  {Kobayashi}}, \bibinfo {author} {\bibfnamefont {M.~G.}\ \bibnamefont
  {House}}, \bibinfo {author} {\bibfnamefont {J.}~\bibnamefont {Slafi}},
  \bibinfo {author} {\bibfnamefont {S.}~\bibnamefont {Barraud}}, \bibinfo
  {author} {\bibfnamefont {R.}~\bibnamefont {Lavieville}}, \bibinfo {author}
  {\bibfnamefont {M.~Y.}\ \bibnamefont {Simmons}}, \ and\ \bibinfo {author}
  {\bibfnamefont {S.}~\bibnamefont {Rogge}},\ }\href@noop {} {\bibfield
  {journal} {\bibinfo  {journal} {arXiv:}\ ,\ \bibinfo {pages} {1703.03538}}
  (\bibinfo {year} {2017})}\BibitemShut {NoStop}%
\bibitem [{\citenamefont {Dirksen}\ \emph {et~al.}(1989)\citenamefont
  {Dirksen}, \citenamefont {Henstra},\ and\ \citenamefont
  {Wenckebach}}]{Dirksen89JPCM}%
  \BibitemOpen
  \bibfield  {author} {\bibinfo {author} {\bibfnamefont {P.}~\bibnamefont
  {Dirksen}}, \bibinfo {author} {\bibfnamefont {A.}~\bibnamefont {Henstra}}, \
  and\ \bibinfo {author} {\bibfnamefont {W.~T.}\ \bibnamefont {Wenckebach}},\
  }\href@noop {} {\bibfield  {journal} {\bibinfo  {journal} {J. Phys.: Condens.
  Matter}\ }\textbf {\bibinfo {volume} {1}},\ \bibinfo {pages} {8535} (\bibinfo
  {year} {1989})}\BibitemShut {NoStop}%
\bibitem [{\citenamefont {Ruskov}\ and\ \citenamefont
  {Tahan}(2013)}]{Ruskov13PRB}%
  \BibitemOpen
  \bibfield  {author} {\bibinfo {author} {\bibfnamefont {R.}~\bibnamefont
  {Ruskov}}\ and\ \bibinfo {author} {\bibfnamefont {C.}~\bibnamefont {Tahan}},\
  }\href@noop {} {\bibfield  {journal} {\bibinfo  {journal} {Phys. Rev. B}\
  }\textbf {\bibinfo {volume} {88}},\ \bibinfo {pages} {064308} (\bibinfo
  {year} {2013})}\BibitemShut {NoStop}%
\bibitem [{\citenamefont {Neubrand}(1978)}]{Neubrand78PSSB}%
  \BibitemOpen
  \bibfield  {author} {\bibinfo {author} {\bibfnamefont {H.}~\bibnamefont
  {Neubrand}},\ }\href@noop {} {\bibfield  {journal} {\bibinfo  {journal}
  {Phys. Status Solidi B}\ }\textbf {\bibinfo {volume} {86}},\ \bibinfo {pages}
  {269} (\bibinfo {year} {1978})}\BibitemShut {NoStop}%
\bibitem [{\citenamefont {Stegner}\ \emph {et~al.}(2010)\citenamefont
  {Stegner}, \citenamefont {Tezuka}, \citenamefont {Andlauer}, \citenamefont
  {Stutzmann}, \citenamefont {Thewalt}, \citenamefont {Brandt},\ and\
  \citenamefont {Itoh}}]{Stegner10PRB}%
  \BibitemOpen
  \bibfield  {author} {\bibinfo {author} {\bibfnamefont {A.~R.}\ \bibnamefont
  {Stegner}}, \bibinfo {author} {\bibfnamefont {H.}~\bibnamefont {Tezuka}},
  \bibinfo {author} {\bibfnamefont {T.}~\bibnamefont {Andlauer}}, \bibinfo
  {author} {\bibfnamefont {M.}~\bibnamefont {Stutzmann}}, \bibinfo {author}
  {\bibfnamefont {M.~L.~W.}\ \bibnamefont {Thewalt}}, \bibinfo {author}
  {\bibfnamefont {M.~S.}\ \bibnamefont {Brandt}}, \ and\ \bibinfo {author}
  {\bibfnamefont {K.~M.}\ \bibnamefont {Itoh}},\ }\href@noop {} {\bibfield
  {journal} {\bibinfo  {journal} {Phys. Rev. B}\ }\textbf {\bibinfo {volume}
  {82}},\ \bibinfo {pages} {115213} (\bibinfo {year} {2010})}\BibitemShut
  {NoStop}%
\bibitem [{\citenamefont {Feher}\ \emph {et~al.}(1960)\citenamefont {Feher},
  \citenamefont {Hensel},\ and\ \citenamefont {Gere}}]{Feher60PRL}%
  \BibitemOpen
  \bibfield  {author} {\bibinfo {author} {\bibfnamefont {G.}~\bibnamefont
  {Feher}}, \bibinfo {author} {\bibfnamefont {J.~C.}\ \bibnamefont {Hensel}}, \
  and\ \bibinfo {author} {\bibfnamefont {E.~A.}\ \bibnamefont {Gere}},\
  }\href@noop {} {\bibfield  {journal} {\bibinfo  {journal} {Phys. Rev. Lett.}\
  }\textbf {\bibinfo {volume} {5}},\ \bibinfo {pages} {309} (\bibinfo {year}
  {1960})}\BibitemShut {NoStop}%
\bibitem [{\citenamefont {Mims}(1968)}]{Mims68PR}%
  \BibitemOpen
  \bibfield  {author} {\bibinfo {author} {\bibfnamefont {W.~B.}\ \bibnamefont
  {Mims}},\ }\href@noop {} {\bibfield  {journal} {\bibinfo  {journal} {Phys.
  Rev.}\ }\textbf {\bibinfo {volume} {168}},\ \bibinfo {pages} {370} (\bibinfo
  {year} {1968})}\BibitemShut {NoStop}%
\bibitem [{\citenamefont {Abadillo-Uriel}\ \emph {et~al.}(2018)\citenamefont
  {Abadillo-Uriel}, \citenamefont {Salfi}, \citenamefont {Hu}, \citenamefont
  {Rogge}, \citenamefont {Calder{\'o}n},\ and\ \citenamefont
  {Culcer}}]{AbadilloUriel18APL}%
  \BibitemOpen
  \bibfield  {author} {\bibinfo {author} {\bibfnamefont {J.~C.}\ \bibnamefont
  {Abadillo-Uriel}}, \bibinfo {author} {\bibfnamefont {J.}~\bibnamefont
  {Salfi}}, \bibinfo {author} {\bibfnamefont {X.}~\bibnamefont {Hu}}, \bibinfo
  {author} {\bibfnamefont {S.}~\bibnamefont {Rogge}}, \bibinfo {author}
  {\bibfnamefont {M.~J.}\ \bibnamefont {Calder{\'o}n}}, \ and\ \bibinfo
  {author} {\bibfnamefont {D.}~\bibnamefont {Culcer}},\ }\href@noop {}
  {\bibfield  {journal} {\bibinfo  {journal} {Appl. Phys. Lett.}\ }\textbf
  {\bibinfo {volume} {113}},\ \bibinfo {pages} {012102} (\bibinfo {year}
  {2018})}\BibitemShut {NoStop}%
\bibitem [{\citenamefont {K\"opf}\ and\ \citenamefont
  {Lassmann}(1992)}]{Kopf92PRL}%
  \BibitemOpen
  \bibfield  {author} {\bibinfo {author} {\bibfnamefont {A.}~\bibnamefont
  {K\"opf}}\ and\ \bibinfo {author} {\bibfnamefont {K.}~\bibnamefont
  {Lassmann}},\ }\href@noop {} {\bibfield  {journal} {\bibinfo  {journal}
  {Phys. Rev. Lett.}\ }\textbf {\bibinfo {volume} {69}},\ \bibinfo {pages}
  {1580} (\bibinfo {year} {1992})}\BibitemShut {NoStop}%
\bibitem [{\citenamefont {Lambert}\ \emph {et~al.}(2018)\citenamefont
  {Lambert}, \citenamefont {Cirio}, \citenamefont {Delbecq}, \citenamefont
  {Allison}, \citenamefont {Marx}, \citenamefont {Tarucha},\ and\ \citenamefont
  {Nori}}]{Lambert18PRB}%
  \BibitemOpen
  \bibfield  {author} {\bibinfo {author} {\bibfnamefont {N.}~\bibnamefont
  {Lambert}}, \bibinfo {author} {\bibfnamefont {M.}~\bibnamefont {Cirio}},
  \bibinfo {author} {\bibfnamefont {M.}~\bibnamefont {Delbecq}}, \bibinfo
  {author} {\bibfnamefont {G.}~\bibnamefont {Allison}}, \bibinfo {author}
  {\bibfnamefont {M.}~\bibnamefont {Marx}}, \bibinfo {author} {\bibfnamefont
  {S.}~\bibnamefont {Tarucha}}, \ and\ \bibinfo {author} {\bibfnamefont
  {F.}~\bibnamefont {Nori}},\ }\href@noop {} {\bibfield  {journal} {\bibinfo
  {journal} {Phys. Rev. B}\ }\textbf {\bibinfo {volume} {97}},\ \bibinfo
  {pages} {125429} (\bibinfo {year} {2018})}\BibitemShut {NoStop}%
\bibitem [{\citenamefont {Bir}\ \emph {et~al.}(1963{\natexlab{a}})\citenamefont
  {Bir}, \citenamefont {Butekov},\ and\ \citenamefont {Pikus}}]{Bir63JPCS}%
  \BibitemOpen
  \bibfield  {author} {\bibinfo {author} {\bibfnamefont {G.}~\bibnamefont
  {Bir}}, \bibinfo {author} {\bibfnamefont {E.}~\bibnamefont {Butekov}}, \ and\
  \bibinfo {author} {\bibfnamefont {G.}~\bibnamefont {Pikus}},\ }\href@noop {}
  {\bibfield  {journal} {\bibinfo  {journal} {Journal of Physics and Chemistry
  of Solids}\ }\textbf {\bibinfo {volume} {24}},\ \bibinfo {pages} {1467}
  (\bibinfo {year} {1963}{\natexlab{a}})}\BibitemShut {NoStop}%
\bibitem [{\citenamefont {Bir}\ \emph {et~al.}(1963{\natexlab{b}})\citenamefont
  {Bir}, \citenamefont {Butikov},\ and\ \citenamefont {Pikus}}]{Bir63JPCS-2}%
  \BibitemOpen
  \bibfield  {author} {\bibinfo {author} {\bibfnamefont {G.}~\bibnamefont
  {Bir}}, \bibinfo {author} {\bibfnamefont {E.}~\bibnamefont {Butikov}}, \ and\
  \bibinfo {author} {\bibfnamefont {G.}~\bibnamefont {Pikus}},\ }\href@noop {}
  {\bibfield  {journal} {\bibinfo  {journal} {Journal of Physics and Chemistry
  of Solids}\ }\textbf {\bibinfo {volume} {24}},\ \bibinfo {pages} {1475}
  (\bibinfo {year} {1963}{\natexlab{b}})}\BibitemShut {NoStop}%
\bibitem [{\citenamefont {Landau}\ and\ \citenamefont
  {Lifshitz}(1981)}]{LandauLifshitzBook}%
  \BibitemOpen
  \bibfield  {author} {\bibinfo {author} {\bibfnamefont {L.~D.}\ \bibnamefont
  {Landau}}\ and\ \bibinfo {author} {\bibfnamefont {E.~M.}\ \bibnamefont
  {Lifshitz}},\ }\href@noop {} {\emph {\bibinfo {title} {Quantum Mechanics:
  Non-Relativistic Theory}}}\ (\bibinfo  {publisher} {Pergamon, London},\
  \bibinfo {year} {1981})\BibitemShut {NoStop}%
\bibitem [{\citenamefont {White}(1973)}]{White73JPDAP}%
  \BibitemOpen
  \bibfield  {author} {\bibinfo {author} {\bibfnamefont {G.~K.}\ \bibnamefont
  {White}},\ }\href@noop {} {\bibfield  {journal} {\bibinfo  {journal} {J.
  Phys. D: Appl. Phys.}\ }\textbf {\bibinfo {volume} {6}},\ \bibinfo {pages}
  {2070} (\bibinfo {year} {1973})}\BibitemShut {NoStop}%
\bibitem [{\citenamefont {Olsen}\ and\ \citenamefont
  {Ettenberg}(1977)}]{Olsen77JAP}%
  \BibitemOpen
  \bibfield  {author} {\bibinfo {author} {\bibfnamefont {G.~H.}\ \bibnamefont
  {Olsen}}\ and\ \bibinfo {author} {\bibfnamefont {M.}~\bibnamefont
  {Ettenberg}},\ }\href@noop {} {\bibfield  {journal} {\bibinfo  {journal} {J.
  Appl. Phys.}\ }\textbf {\bibinfo {volume} {48}},\ \bibinfo {pages} {2543}
  (\bibinfo {year} {1977})}\BibitemShut {NoStop}%
\end{thebibliography}%

\newpage


\section*{Supplemental Material}
\noindent\textbf{Acceptor Hamiltonian.}
Si:B holes couple to uniform magnetic, electric and elastic fields as described in Refs.~\onlinecite{Bir63JPCS} and \onlinecite{Bir63JPCS-2}.
The coupling Hamiltonian to the fields is expressed as follows:
\begin{eqnarray}
\label{Hamiltonian}
\hat{H}^\prime &=& \hat{H}_B^\prime + \hat{H}_{E,ion}^\prime + \hat{H}_E^\prime + \hat{H}_\varepsilon^\prime,\label{Hamiltonian}\\
\hat{H}_B^\prime &=& \mu_B\sum_{i=x,y,z} \left(g_1^\prime \hat{J}_i+g_2^\prime{\hat{J}_i}^3\right)B_{0,i},\nonumber\\
\hat{H}_{E,\rm{ion}}^\prime &=& \frac{2p}{\sqrt{3}}\left( E_x\hat{Q}_{yz} + E_y\hat{Q}_{zx} + E_z\hat{Q}_{xy} \right),\nonumber\\
\hat{H}_\varepsilon^\prime &=& b^\prime\sum_{i=x,y,z}\varepsilon_{ii}\hat{Q}_{ii} + \frac{2d^\prime}{\sqrt{3}}\left( \varepsilon_{xy}\hat{Q}_{xy} + \varepsilon_{yz}\hat{Q}_{yz} + \varepsilon_{zx}\hat{Q}_{zx} \right),\nonumber\\
\hat{H}_E^\prime &=& b\sum_{i=x,y,z}E_{i}^2\hat{Q}_{ii} + \frac{2d}{\sqrt{3}}\left( E_xE_y\hat{Q}_{xy} + E_yE_z\hat{Q}_{yz} + E_zE_x\hat{Q}_{zx} \right),\nonumber
\end{eqnarray}
where $x$, $y$ and $z$ axes correspond to the $[100]$, $[010]$ and $[001]$ axes of silicon crystal, respectively, and  $\hat{I}$ is the identity operator. 
Quadrupole operators $\hat{Q}_{ij}$ ($i,j=x,y,z$) for $J=\tfrac{3}{2}$ are expressed by angular momentum operators $\hat{J}_i$ as $\hat{Q}_{ij}=\tfrac{1}{2}\{\hat{J}_i,\hat{J}_j\}-\tfrac{5}{4}\delta_{ij}\hat{I}$, where $\{\hat{J}_i,\hat{J}_j\}=\hat{J}_i \hat{J}_j+\hat{J}_j \hat{J}_i$ and $\delta_{ij}$ is Kronecker's delta.
$E_i$ are electric field and $\varepsilon_{ij}$ are normal (shear) strain for $i=j$ ($i\neq j$).
Notice that all interactions with electrical and elastic degrees of freedom are proportional to $\hat{Q}_{ij}$ \cite{LandauLifshitzBook,Winkler04PRB} and as such they couple the light and heavy holes.
While $H_\varepsilon^\prime$ generally includes a hydrostatic term like $a\hat{I}$, it does not cause any relative energy change of $J=\tfrac{3}{2}$ hole states and thus is dropped off here.

For the coefficients, we use g-factors $g_1^\prime=-1.07$ and $g_2^\prime=-0.03$ \cite{Neubrand78PSSB}, linear electric-field coupling coefficient $p=0.26~\text{Debye}$ \cite{Kopf92PRL} and deformation potentials $b^\prime=-1.42~\text{eV}$ and $d^\prime=-3.7~\text{eV}$ \cite{Neubrand78PSSB}.
$b$ and $d$ are cubic electric-field coupling coefficients. 
While there is no experimental information on the values of $b$ and $d$, we estimate them at $\sim -3~\text{Debye/(MV/m)}$ and $\sim -5~\text{Debye/(MV/m)}$, respectively, based on the effective mass approach in Ref.~\onlinecite{Bir63JPCS-2}.

By numerically diagonalising $H^\prime$ with certain sets of $\overrightarrow{E}$ and $\varepsilon_{ij}$ as a function of  $\overrightarrow{B_0}$, we obtain eigenenergy spectra as shown in Figs.~\ref{Concept}a and b: $\varepsilon_{xx}=\varepsilon_{yy}=\varepsilon_{zz}=0$ and $\overrightarrow{B_0}\parallel[110]$ for Fig.~\ref{Concept}b, and $\varepsilon_{xx}=\varepsilon_{yy}=0.02~\%$, $\varepsilon_{zz}=-0.0156~\%$ and $\overrightarrow{B_0}\parallel[110]$ for Fig.~\ref{Concept}c. ($|\overrightarrow{E}|$ and $\varepsilon_{ij}$ ($i\neq j$) are kept zero.)

\noindent\textbf{Experimental setup.}
Both of the mechanically relaxed and strained samples are prepared from pieces of the same boron-doped $^{28}$Si wafer with dimensions of 4.0 mm $\times$ 3.5 mm in area and 500 $\mu$m in thickness, $^{28}$Si-purity of 99.99+ \% and boron concentration $n_{\text{B}}$ of $1.0$--$1.5\times10^{15}~\text{cm}^{-3}$.
A diced piece of this crystal is used for the relaxed sample.
To prepare the strained sample, another piece is thinned down to $50~\mu\text{m}$ and glued to a fused silica chip (5.0 mm $\times$ 5.0 mm in area and 1 mm in thickness) by two-component epoxy adhesive (Fig.~\ref{Spectra}a).
While both the $^{28}$Si and fused silica chips in this stack are mechanically relaxed at room temperature, biaxial tensile strain is applied to the $^{28}$Si chip at low temperature owing to the difference in thermal expansion coefficient of these two materials \cite{White73JPDAP}. 
The magnitude of the strain applied to the strained sample is discussed in the section \textit{Strain analyses}.

\begin{figure*}
[ptb]
\begin{center}
\includegraphics{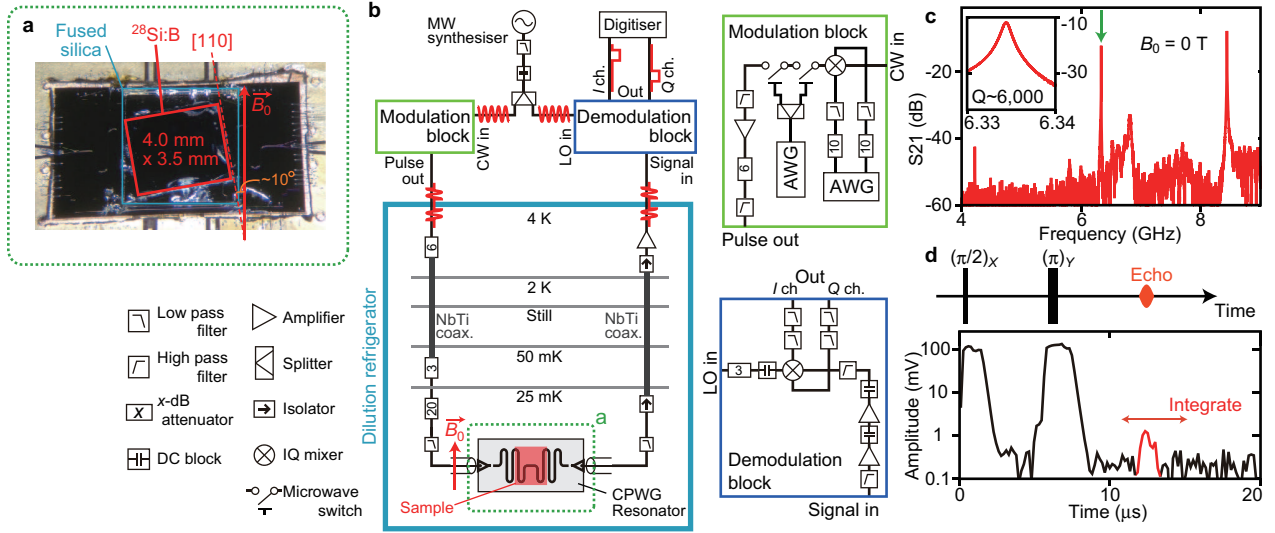}
\caption{\textbf{a,} Photo image of the strained $^{28}$Si:B sample on the niobium coplanar waveguide resonator. 
\textbf{b,} Schematic figure of the experimental setup. 
\textbf{c,} Microwave transmission spectrum of the experimental setup measured at zero magnetic field by a network analyser. The sold arrow indicates the resonance mode used for spin-echo experiments. Inset: Detailed transmission spectrum around the resonance used for spin-echo experiments. 
\textbf{d,} Typical output signal for input of a Hahn-echo pulse sequence in time domain. 
}%
\label{Setup}
\end{center}
\end{figure*}

To perform EPR spectroscopy at millikelvin temperature, we use a superconducting coplanar waveguide cavity with a small mode volume.
The cavity is fabricated from 100-nm-thin niobium film and consists of a 20-$\mu\text{m}$-wide centre conductor and 12-$\mu\text{m}$-wide separations between the centre conductor and ground plates, which yield the characteristic impedance of 50 $\Omega$ on 400-$\mu\text{m}$-thick highly resistive silicon substrate. 
Each sample is mounted to the cavity in an independent experimental run to avoid overlapping signals from different samples.
To couple to the cavity modes, the samples are closely fitted to the cavity so that the polished silicon surface faces the cavity structure and fixed by GE varnish (Extended Data Fig.~\ref{Setup}a).

A cavity chip together with a sample is mounted to the lowest temperature stage of a dilution refrigerator with EPR experimental set up as shown in Extended Data Fig.~\ref{Setup}b.
Extended Data Fig.~\ref{Setup}c shows a typical transmission spectrum through the refrigerator with the cavity cooled down to $\sim25~\text{mK}$ at zero magnetic field.
Cavity resonance modes appear at each $\sim2.1~\text{GHz}$.
For EPR experiments, we use the cavity mode at $\sim6.3~\text{GHz}$ (green arrow), which has a linewidth of $\sim1~\text{MHz}$ and a quality factor of $\sim6,000$.
The magnetic field $B_0$ is applied parallel to the resonator surface to maintain the resonator's quality factor, which is more than 3,000 at the magnetic fields used for experiments.
The [110] axis of the samples is nearly aligned to the magnetic field.
For the strained sample, misalignment of the magnetic field with respect to the [110] axis is approximately $10~\text{deg}.$ (Extended Data Fig.~\ref{Setup}a).
In the spin-echo measurements, output from the refrigerator is demodulated into quadrature signals $I(t)$ and $Q(t)$, and recorded by a digitiser.
Spins are manipulated by microwave pulses modulated by the modulation block containing microwave switches and an IQ mixer controlled by arbitrary waveform generators.
$(\pi/2)_X$ and $(\pi)_Y$ pulses used in this work are microwave modulated by $300$-$\text{ns}$-long square pulse without phase shift and $600$-$\text{ns}$-long square pulse with phase shift of $\pi/2$, respectively.
Extended Data Fig.~\ref{Setup}d displays a typical output signal in the form of time-domain amplitude $V(t)=\sqrt{I(t)^2+Q(t)^2}$ after a standard Hahn-echo pulse sequence consisting of $(\pi/2)_X$ and $(\pi)_Y$ pulses and $\tau=6~\mu\text{s}$.
A spin-echo signal emphasised by red lines appears just after large pulsed signals corresponding to the $(\pi/2)_X$ and $(\pi)_Y$ pulses, assuring that we can detect spin signal by using this experimental setup.

\noindent\textbf{Strain analyses.}
A previous report shows that bulk silicon (silica) contracts by $0.021~\%$ ($\sim-0.002~\%$) for each direction when it is cooled down from 293 K to 4 K \cite{White73JPDAP}. 
This thermal expansion mismatch induces biaxial tensile strain of up to $\sim0.023~\%$ in the silicon layer of the stack structure used for the strained sample.
Strain in the actual sample is less than this estimation and also distributed owing to the finite thickness of the silicon layer.

\begin{figure}
[ptb]
\begin{center}
\includegraphics{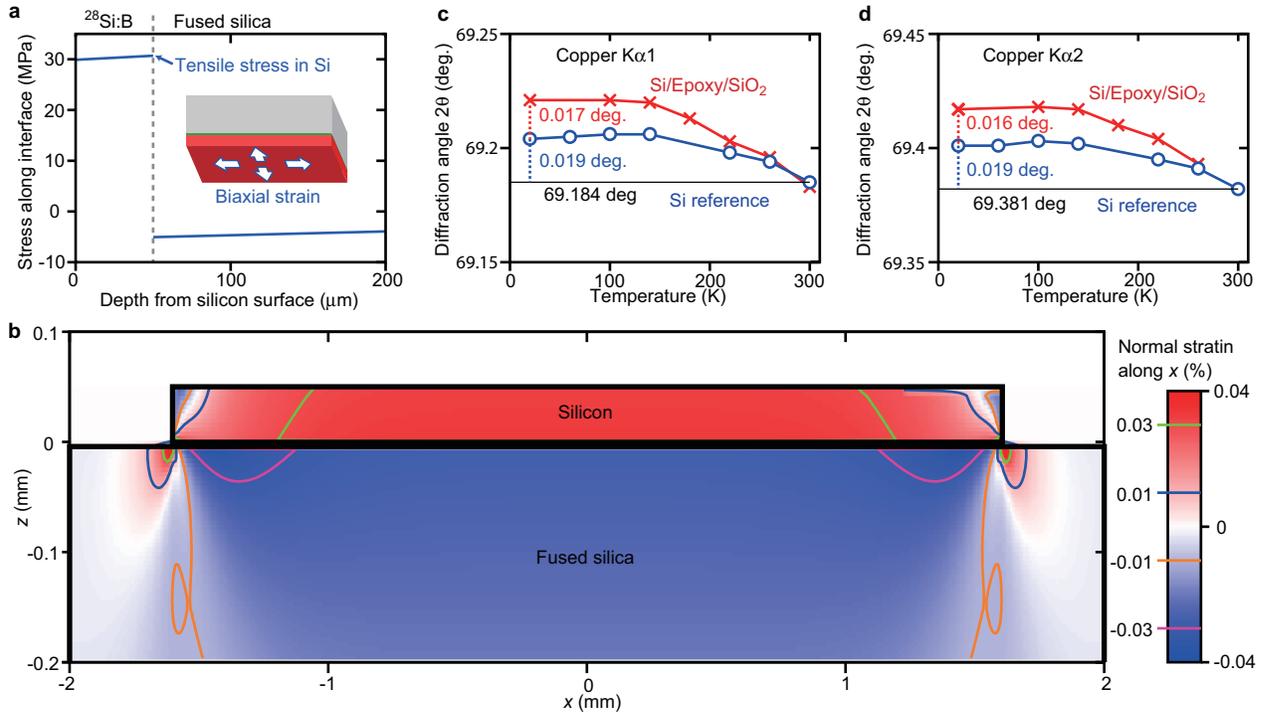}
\caption{\textbf{a,} Theoretical analysis of the stress distribution.
\textbf{b,} Numerical analysis of the strain ($\varepsilon_{xx}$) distribution.
\textbf{c,d,} Temperature dependence of X-ray diffraction angle of Si(004) measured by using the copper $\text{K}\alpha1$ emission line (\textbf{c}) and the $\text{K}\alpha2$ emission line (\textbf{d}).
}%
\label{Strain}
\end{center}
\end{figure}

Based on a more sophisticated calculation of stress in a stack structure \cite{Olsen77JAP}, we expect biaxial tensile stress of $\sim 30~\text{MPa}$ is induced to the silicon layer (Extended Data Fig.~\ref{Strain}a). 
Assuming Young's modulus $130~\text{GPa}$ and Poisson's ratio $0.28$ along the [100] axis in silicon, we find that tensile biaxial strain of $\sim0.017~\%$ is induced to the silicon chip. 
Numerical simulation of the strain distribution in the stack structure also shows biaxial tensile strain of $0.032~\%$ at the centre of the sample (Extended Data Fig.~\ref{Strain}b).
This simulation also implies that, while strain is uniform around the centre of the silicon layer, decreases and eventually turns to compressive strain with approaching the sample edge.
This strain variation induces difference between strain along $x$ and $y$ axis and thus could result in the distribution in $g^\ast$ value of boron spins in the strained sample (see the section \textit{Land\'e g-factor distribution induced by strain anisotropy}).

We also perform X-ray diffraction analysis at low temperature by using an identical stack structure by using $^\text{nat}$Si wafer instead of $^{28}$Si.
While the silicon (400) diffraction angles taken from the stack structure and a reference silicon sample well coincide at room temperature, they gradually move apart as temperature decreases (Extended Data Figs.~\ref{Strain}c and d).
This indicates that the silicon layer in the stack structure is contracted perpendicularly to the surface more than normal thermal contraction observed in the reference sample.
This extra contraction in the stack structure is associated with stretch along the silicon surface as expected.
From these measurements, we estimate biaxial strain in the silicon layer of the stack structure at $0.028~\%$ tensile which is consistent with the expectation from the difference in thermal contraction.

\begin{figure*}
[ptb]
\begin{center}
\includegraphics{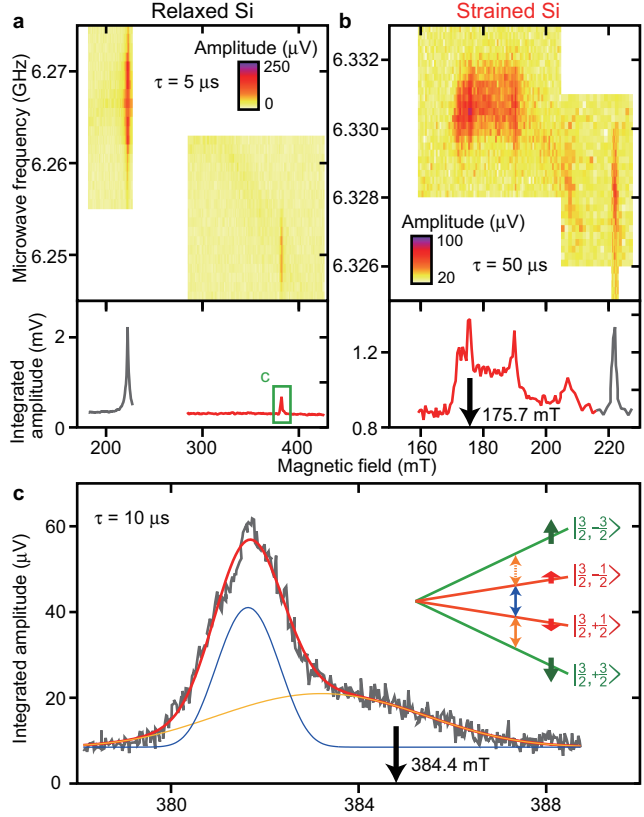}
\caption{\textbf{a,b,} Spin-echo spectra as a function of magnetic field and microwave frequency (top panels) for the relaxed (\textbf{a}) and strained samples (\textbf{b}). Each data point is obtained by integrating time domain signals as shown in Fig.~\ref{Setup}d. For comparison, spin-echo spectra same as Figs.~\ref{Spectra}b and c, obtained by integrating data in the top panels along the microwave frequency axis, are shown in the bottom panels.
\textbf{c,} Detailed spin-echo spectra in the box in the bottom panel of a measured with $\tau = 10~\mu\text{s}$. The black arrows in the c and the bottom panel of b indicate the magnetic field used to measure $T_2$ and $T_1$ in the relaxed (384.4 mT) and strained samples (175.7 mT), respectively. The red curve shows the fitting function, which is composed of two Gaussian functions shown by the blue and orange curves. Inset: Level spectrum of light-hole and heavy-hole states in relaxed Si:B. Blue and orange arrows show transitions that provide spin-echo spectra fitted by the same coloured curves in the main panel.
}%
\label{DetailedSpectra}
\end{center}
\end{figure*}

\noindent\textbf{Detail of spin-echo measurements.}
In Hahn echo measurements, recorded signals $I(t)$ and $Q(t)$ are converted to an time-domain amplitude $V(t)$ and subsequently integrated over a time span around the echo signal as shown by the red arrow in Extended Data Fig.~\ref{Setup}d.
Data plotted in Fig.~\ref{T1T2} are taken by repeat of this process with changing $\tau$ or $t^\prime$ at certain $|\overrightarrow{B_0}|$ and $\omega_{\text{MW}}/2\pi$.
For Fig.~\ref{Spectra}, we first take integrated amplitude signals with changing not only $|\overrightarrow{B_0}|$ but also $\omega_{\text{MW}}/2\pi$, since the resonant frequency of the superconducting resonator is slightly shifted as $|\overrightarrow{B_0}|$ is changed.
Extended Data Figs.~\ref{DetailedSpectra}a and b display typical integrated amplitude spectra taken with sweeping $|\overrightarrow{B_0}|$ and $\omega_{\text{MW}}/2\pi$ (top panels), showing that microwave frequency providing spin-echo signal changes gradually with changing magnetic field.
By integrating these data over $\omega_{\text{MW}}/2\pi$, we finally obtain spin-echo spectra as a function of $|\overrightarrow{B_0}|$ (bottom panels) as shown in Fig.~\ref{Spectra}.
Note that, experiments for the relaxed and strained samples are carried out in different experimental runs, the cavity resonance and thus spin echo signal appear at slightly different frequencies between the relaxed and strained samples. 
In CPMG measurements, the time-domain echo signal appears after each of $(\pi)_Y$ and $(-\pi)_Y$ pulses as schematised in the inset of Fig.~\ref{CPMG}.
By integrating each of them over time independently, we obtain the CPMG-echo decay signal as a function of elapsed time after the first $(\pi/2)_X$ pulse. 
Since, in general, spin-echo signals that appear after $(\pi)_Y$ and $(-\pi)_Y$ pulses are not equivalent, we plot echo-signals accompanying $(\pi)_Y$ pulses only in Fig.~\ref{CPMG} and use them for fitting.

Extended Data Fig.~\ref{DetailedSpectra}c shows the fine structure of the boron spin-echo spectrum in the relaxed sample. 
The spectrum consists of a sharp line and backgrounding broad line rather than simple single peak, indicating that two different spin transitions occur at almost the same magnetic field.
This observation is reasonable for the relaxed sample because the $|m_J|=\tfrac{3}{2}\leftrightarrow\tfrac{1}{2}$ and $m_J=+\tfrac{1}{2}\leftrightarrow-\tfrac{1}{2}$ transitions (orange and blue arrows in the inset) have almost same transition frequency.
As reported in Refs.~\onlinecite{Feher60PRL,Neubrand78PSSB,Stegner10PRB}, $|m_J|=\tfrac{3}{2}\leftrightarrow\tfrac{1}{2}$ transition linewidth is inhomogeneously broadened by random strain in the crystal more strongly than the $m_J=+\tfrac{1}{2}\leftrightarrow-\tfrac{1}{2}$ transition.
Hence, we attribute the broad spin-echo peak to the $m_J=+\tfrac{3}{2}\leftrightarrow+\tfrac{1}{2}$ and $m_J=-\tfrac{1}{2}\leftrightarrow-\tfrac{3}{2}$ transitions, while the sharp peak to the $m_J=+\tfrac{1}{2}\leftrightarrow-\tfrac{1}{2}$ transition.
We fit a sum of two Gaussian functions $f_1(|\overrightarrow{B_0}|)+f_2(|\overrightarrow{B_0}|)$ (red curve), where  $f_n(|\overrightarrow{B_0}|)=A\text{exp}\{-(|\overrightarrow{B_0}|-B_{c,n})^2/{\delta_{n}}^2\}$ ($n=1,2$), to the data, obtaining a linewidth $\delta_{n}$ of $3.7~\text{mT}$ ($1.0~\text{mT}$) for the broader (sharper) spin-echo peak as shown by the orange (blue) curve.
These linewidths well coincide with previously measurements in relaxed $^{28}$Si:B \cite{Stegner10PRB}.
We note that, while $^{28}$Si:B resonance lines are reported to be well fitted by Lorentzian functions \cite{Stegner10PRB}, Lorentzian fit of our peak shape implies unrealistic thermal population of hole spin states and thus we use Gaussian functions for fitting.

\begin{figure}
[ptb]
\begin{center}
\includegraphics{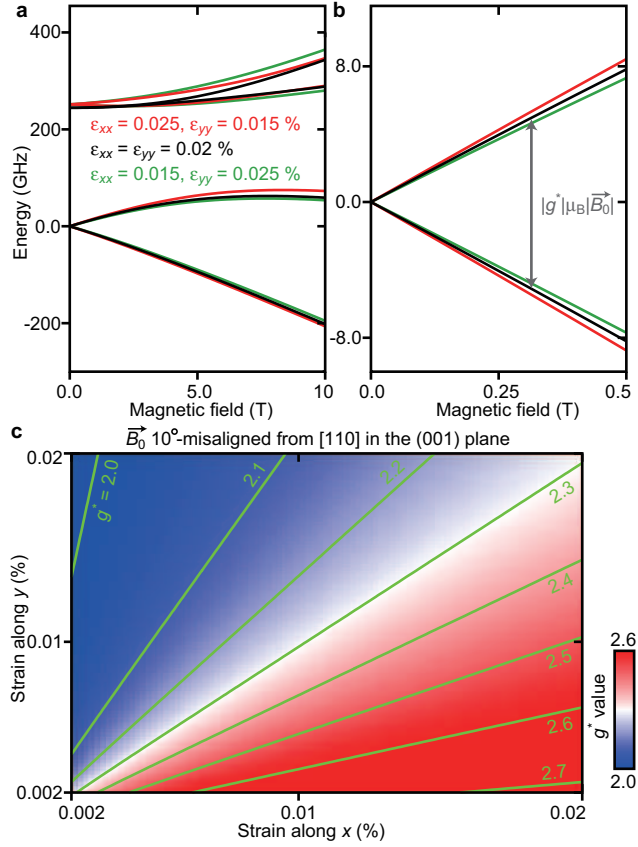}
\caption{\textbf{a,b,} Eigenenergy spectra as a function of magnetic field for three different sets of $\varepsilon_{xx}$ and $\varepsilon_{yy}$ with same $\varepsilon_{xx}+\varepsilon_{yy}$: $(\varepsilon_{xx}, \varepsilon_{yy}) = (0.02~\%, 0.02~\%)$ (black), $(0.025~\%, 0.015~\%)$ (red) and $(0.015~\%, 0.025~\%)$ (green). We estimate $g^\ast$ value from the magnetic field dependence of level splitting as shown in b.
\textbf{c,} $\varepsilon_{xx}$ and $\varepsilon_{yy}$ dependence of $g^\ast$ for magnetic field of 175.7 mT misaligned by 10 deg. from the [110] axis in the (001) plane.
}%
\label{Distribution}
\end{center}
\end{figure}

\noindent\textbf{Land\'e g-factor distribution induced by strain anisotropy.}
As implied by Extended Data Fig.~\ref{Strain}b, strain in the stack structure for the strained sample is not perfectly biaxial ($\varepsilon_{xx}=\varepsilon_{yy}$) but strain along the [100] and [010] axes can be different ($\varepsilon_{xx}\neq\varepsilon_{yy}$). 
Extended Data Fig.~\ref{Distribution}a shows eigenenergy spectra of an acceptor-bound hole for three different sets of $\varepsilon_{xx}$ and $\varepsilon_{yy}$ with same $\varepsilon_{xx}+\varepsilon_{yy}$: $(\varepsilon_{xx}, \varepsilon_{yy}) = (0.02~\%, 0.02~\%)$, $(0.025~\%, 0.015~\%)$ and $(0.015~\%, 0.025~\%)$.
A marked difference appears in the magnetic field dependence, while energy splitting at zero magnetic field shows only a little change.
The magnetic field dependence in the low magnetic field regime (Extended Data Fig.~\ref{Distribution}b) is relevant to this work, characterised by the effective g-factor $g^\ast$.
We calculate $g^*$ value from such energy spectra and plot them as a function of $\varepsilon_{xx}$ and $\varepsilon_{yy}$ as shown in Extended Data Fig.~\ref{Distribution}c.
Here we assume $\overrightarrow{B_0}$ is not perfectly aligned to the [110] axis but misaligned by $\sim10$ degrees. 
The $g^*$ value is changed by $\sim0.1$ when strain anisotropy $(\varepsilon_{xx}-\varepsilon_{yy})/2(\varepsilon_{xx}+\varepsilon_{yy})$ is $\sim 20~\%$.
Since Extended Data Fig.~\ref{Strain}b implies that strong strain anisotropy appears near the edges of the silicon layer, the observed distribution in $g^*$ value is presumably attributed to strain anisotropy in the real sample.

\begin{figure}
[ptb]
\begin{center}
\includegraphics{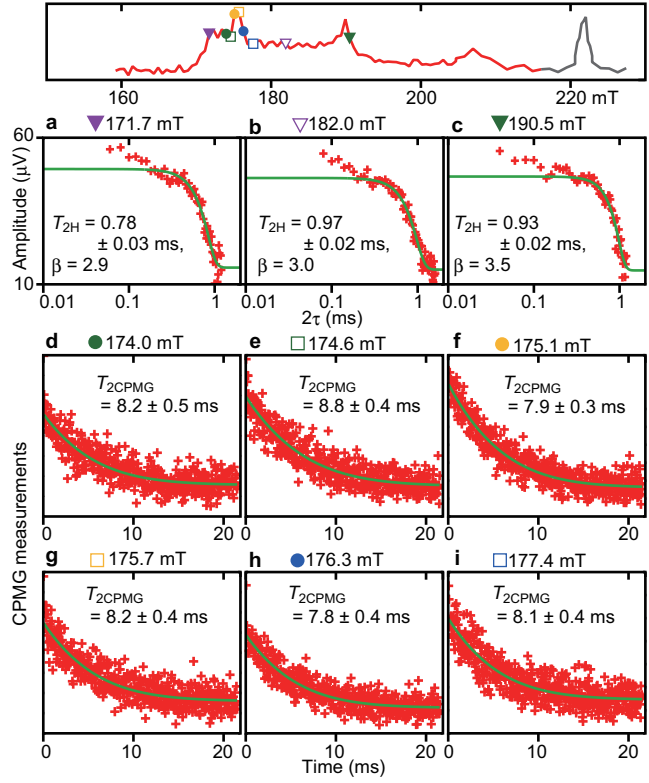}
\caption{\textbf{a-c,} Hahn-echo decay measurements at three different magnetic field.
\textbf{d-i,} CPMG-echo decay measurements at six different magnetic field.
Magnetic field used for these measurements are indicated by symbol corresponding to each of them in a spin-echo spectrum in the top panel.
}%
\label{T2s}
\end{center}
\end{figure}

\noindent\textbf{Magnetic field used to measure data in Figs.~\ref{T1T2} and \ref{CPMG}.}
While the $|m_J|=\tfrac{3}{2}\leftrightarrow\tfrac{1}{2}$ transitions have electric dipole moment as mentioned in the main text, the $m_J=+\tfrac{1}{2}\leftrightarrow-\tfrac{1}{2}$ transition does not in relaxed silicon.
To discuss effect of electric dipole moment in coherence, we need to address the $m_J=+\tfrac{3}{2}\leftrightarrow+\tfrac{1}{2}$ transition without exciting the $m_J=+\tfrac{1}{2}\leftrightarrow-\tfrac{1}{2}$ transition. 
By measuring at $|\overrightarrow{B_0}|=384.4~\text{mT}$ (black arrow in Extended Data Fig.~\ref{DetailedSpectra}c) and $\omega_{\text{MW}}/2\pi=6.255~\text{GHz}$, we eliminate signal attributed to the $m_J=+\tfrac{1}{2}\leftrightarrow-\tfrac{1}{2}$ transition from spin-echo decay measurements.
In addition the spin temperature is estimated at $\sim300~\text{mK}$ from ratio of the Hahn-echo intensity between the $|m_J|=\tfrac{3}{2}\leftrightarrow\tfrac{1}{2}$ transitions and the $m_J=+\tfrac{1}{2}\leftrightarrow-\tfrac{1}{2}$ transition.
This assures that the thermal population of the $\left|\tfrac{3}{2},-\tfrac{1}{2}\right\rangle$ state is quite small and thus the $m_J=-\tfrac{1}{2}\leftrightarrow-\tfrac{3}{2}$ transition does not contribute to the observed Hahn-echo signal.
In contrast, spin-echo decay in the strained sample does not show clear dependence in magnetic field (Extended Data Figs.~\ref{T2s}a-c).
We also confirmed that the CPMG decay is also not influenced to a peak-like substructure around $|\overrightarrow{B_0}|=175.7~\text{mT}$ in the spin echo spectrum as shown in Extended Data Figs.~\ref{T2s}d-i.
While $T_{2\text{CPMG}}$'s obtained from these measurements are slightly shorter than presented in the main text, this is presumably attributed to the signal to noise ratio in Extended Data Figs.~\ref{T2s}d-i lower than Fig.~\ref{CPMG}.

\noindent\textbf{Electric dipole moment and coupling to phonons.}
The effective Hamiltonian that describes the response of the qubit to electric fields $\overrightarrow{E}$ is
\begin{eqnarray}
\hat{H}_{\rm qbt}&=&\tfrac{1}{2}\hbar(\omega_0 + \omega)\hat{\sigma}_Z + \sum_{\alpha=X,Y}\hbar\Omega_\alpha\hat{\sigma}_\alpha,\nonumber\\
\hbar\omega &\equiv& \overrightarrow{\chi}(\overrightarrow{E})\cdot\overrightarrow{E} = \sum_{i=x,y,z}\left(\chi_{i} + \sum_{j=x,y,z}\chi_{ij}E_{j}\right)E_{i},\label{Dipole1}\\
\hbar\Omega_\alpha &\equiv& \overrightarrow{v_\alpha}(\overrightarrow{E})\cdot\overrightarrow{E} = \sum_{i=x,y,z}\left(v_{\alpha i} + \sum_{j=x,y,z}v_{\alpha ij}E_{j}\right)E_{i}.\label{Dipole2}
\end{eqnarray}
Here, $\omega_0$ is the qubit Larmor precession frequency without electric fields. 
$\omega$ is the change in Larmor frequency with electric fields.
Fluctuations of $\omega$, $\delta\omega=\overrightarrow{\chi}(\overrightarrow{E})\cdot\delta\overrightarrow{E}$, induced by electric field fluctuations $\delta\overrightarrow{E}$ cause decoherence.
$\Omega_\alpha$ is the frequency of nutation around the $\alpha =X,Y$ axes, enabling quantum manipulations of the qubit.
The effective electric dipole moment of the qubit is split to longitudinal component $\overrightarrow{\chi}(\overrightarrow{E})$ and transverse component $\overrightarrow{v_\alpha}(\overrightarrow{E})$ in accordance with the effect in $\omega$ and $\Omega_\alpha$, respectively.
We note that the $XYZ$ coordinate system, in general, coincides neither the real-space $xyz$ coordinate nor the simple rotating frame coordinate commonly used to discuss electron spin resonance. 
Finite strain deviates the spin precession trajectory of holes from the plane perpendicular to the magnetic field, requiring complicated rotations to obtain a frame where the hole spin is static.

We have calculated $\chi_{i}$, $\chi_{ij}$, $v_{\alpha i}$ and $v_{\alpha ij}$ in Eqs.~(\ref{Dipole1}) and (\ref{Dipole2}) for the charge-like subsystem $\{\left|\tfrac{3}{2},+\tfrac{3}{2}\right\rangle, \left|\tfrac{3}{2},+\tfrac{1}{2}\right\rangle\}$ where $\Delta=0$ and for the generalised spin $\{\left|\tfrac{3}{2},+\tfrac{1}{2}\right\rangle, \left|\tfrac{3}{2},-\tfrac{1}{2}\right\rangle\}$ where $\Delta>\hbar\omega_0$ using the Schrieffer--Wolff transformation described in ref.~\onlinecite{Salfi16PRL}.

For relaxed Si:B ($\Delta=0$) subjected to a magnetic field in the (001) plane, the following non-zero electric dipole matrix elements are obtained by directly evaluating elements of Eq.~\ref{Hamiltonian}: 
\begin{align}
&\chi_{z}E_z=\sqrt{3}\sin(2\theta_0)\Big(pE_z\Big), &\chi_{xy}E_{x}E_y=\sqrt{3}\sin(2\theta_0)\Big(dE_xE_y\Big),\nonumber\\ 
& &\chi_{xx}E_x^2=\Big(-3\cos(2\theta_0)-1\Big)\frac{1}{2}bE_x^2,\nonumber\\
& &\chi_{yy}E_y^2=\Big(+3\cos(2\theta_0)-1\Big)\frac{1}{2}bE_y^2,\nonumber\\
& &\chi_{zz}E_z^2=-bE_z^2,\nonumber\\
&v_{Xx}E_x=\sin(\theta_0)\Big(pE_x\Big), &v_{Xyz}E_yE_z=\sin(\theta_0)\Big(dE_yE_z\Big),\nonumber\\
&v_{Xy}E_y=\cos(\theta_0)\Big(pE_y\Big), &v_{Xxz}E_xE_z=\cos(\theta_0)\Big(dE_xE_z\Big),\nonumber\\
&v_{Yz}E_z=\cos(2\theta_0)\Big(pE_z\Big),&v_{Yxy}E_xE_y=\cos(2\theta_0)\Big(dE_xE_y\Big),\nonumber\\
& &v_{Yxx}E_x^2=\frac{\sqrt{3}}{2}\sin(2\theta_0)\Big(bE_x^2\Big),\nonumber\\
& &v_{Yyy}E_y^2=-\frac{\sqrt{3}}{2}\sin(2\theta_0)\Big(bE_y^2\Big).
\end{align}
Here $\theta_0$ is the angle of magnetic field to the [100] direction. 
$p$ is the $\text{T}_{\rm d}$ symmetry linear coupling to electric fields and $d$ is the cubic symmetry second-order couplings to electric fields, which both introduce spin-orbit coupling and are equivalent to those in the original acceptor Hamiltonian $\hat{H^\prime}$ defined by Eq.~(\ref{Hamiltonian}).

The longitudinal relaxation rate ${T_1}^{-1}$ is obtained by using Fermi's golden rule together with the Bir--Pikus deformation potential $b^\prime$ and $d^\prime$ \cite{Bir63JPCS,Bir63JPCS-2,Neubrand78PSSB}
\begin{equation}
{T_1}^{-1}=\frac{(\hbar\omega_0)^3}{20\pi\rho\hbar^4}\Big[{b^\prime}^2\sin^2(2\theta_0)\Big(\frac{2}{v_l^5}+\frac{3}{v_t^5}\Big)+{d^\prime}^2(1+\cos^2(2\theta_0))\Big(\frac{2}{3v_l^5}+\frac{1}{v_t^5}\Big)\Big],\nonumber
\end{equation}
while for $\theta_0=\pi/4$ as in the experiment we have
\begin{equation}
{T_1}^{-1}=\frac{(\hbar\omega_0)^3}{20\pi\rho\hbar^4}\Big[{b^\prime}^2\Big(\frac{2}{v_l^5}+\frac{3}{v_t^5}\Big)+{d^\prime}^2\Big(\frac{2}{3v_l^5}+\frac{1}{v_t^5}\Big)\Big].
\label{T1r}
\end{equation}
Here, $\rho=2990~\text{kg}/\text{m}^3$ is the mass density of silicon and $v_l=8.99\times10^{3}~\text{m/s}$ ($v_t=v_l/1.7$) is the longitudinal (transverse) speed of sound in silicon crystal.
When the gap $\Delta$ dominates $\hbar\omega_0$ we obtain the following non-zero electric dipole matrix elements to lowest order in $\hbar\omega_0/2\Delta$ using a Schrieffer--Wolff transformation: 
\begin{align}
\chi_{z}E_z=&\frac{\sqrt{3}\hbar\omega_0}{2\Delta}\sin(2\theta_0)\Big(pE_z\Big), &\chi_{xy}E_{x}E_y=\frac{\sqrt{3}\hbar\omega_0}{2\Delta}\sin(2\theta_0)\Big(dE_xE_y\Big),\nonumber\\ 
& &\chi_{xx}E_x^2=\frac{\sqrt{3}\hbar\omega_0}{2\Delta}\cos(2\theta_0)\Big(-\sqrt{3}bE_x^2\Big),\nonumber\\
& &\chi_{yy}E_{y}^2=\frac{\sqrt{3}\hbar\omega_0}{2\Delta}\cos(2\theta_0)\Big(\sqrt{3}bE_y^2\Big),\nonumber\\
v_{Xx}E_x=&\frac{\sqrt{3}\hbar\omega_0}{2\Delta}\sin(\theta_0)\Big(pE_x\Big), &v_{Xyz}E_yE_z=\frac{\sqrt{3}\hbar\omega_0}{2\Delta}\sin(\theta_0)\Big(dE_yE_z\Big),\nonumber\\
v_{Xy}E_y=&\frac{\sqrt{3}\hbar\omega_0}{2\Delta}\cos(\theta_0)\Big(pE_y\Big),&v_{Xxz}E_xE_z=\frac{\sqrt{3}\hbar\omega_0}{2\Delta}\cos(\theta_0)\Big(dE_xE_z\Big),\nonumber\\
v_{Yz}E_z=&\frac{\sqrt{3}\hbar\omega_0}{2\Delta}\cos(2\theta_0)\Big(pE_z\Big),&v_{Yxy}E_xE_y=\frac{\sqrt{3}\hbar\omega_0}{2\Delta}\cos(2\theta_0)\Big(dE_xE_y\Big),\nonumber\\
& &v_{Yxx}E_x^2=\frac{\sqrt{3}\hbar\omega_0}{2\Delta}\sin(2\theta_0)\Big(\frac{\sqrt{3}}{2}bE_x^2\Big),\nonumber\\
& &v_{Yyy}E_y^2=\frac{\sqrt{3}\hbar\omega_0}{2\Delta}\sin(2\theta_0)\Big(-\frac{\sqrt{3}}{2}bE_y^2\Big).
\end{align}
All other terms are zero to linear order in $\hbar\omega_0/\Delta$. 
Please note that the $bJ_z^2E_z^2$ term in $H'_E$ does not contribute to the longitudinal dipole that causes decoherence, that is $\chi_{zz}E_z^2=0$, in contrast to the relaxed case. 
This fact holds to any order $n>1$ in $(\hbar\omega_0/\Delta)^n$ for the generalised spin two-level system because $J_z^2$ is diagonal in the $J=\tfrac{3}{2}$ basis and it is the identity matrix in the generalised spin subspace. Also note that this means that logic gates implementing a periodic transverse coupling $\overrightarrow{v_\alpha}(t)$ using a periodic modulation of $E_z(t)$ would not cause a shift in Larmor frequency during the gate, only an oscillating component that could easily be made to average to zero with no effect on qubit coherence. This is convenient since a sinusoidal control field $E_z(t)$ could be implemented with a top gate to control the transverse dipole.

The longitudinal relaxation rate ${T_1}^{-1}$ for $\Delta > \hbar\omega_0$ is calculated by projecting the elastic interactions into the qubit subspace by a Schrieffer--Wolff transformation. We obtain the result:
\begin{eqnarray}
{T_1}^{-1}=\frac{(\hbar\omega_0)^3}{20\pi\rho\hbar^4}\Big(\frac{\hbar\omega_0}{\Delta}\Big)^2\Big[& &\frac{{b^\prime}^2(72\sin^2(2\theta_0)+9)}{128}\Big(\frac{4}{3v_l^5}+\frac{2}{v_t^5}\Big)\nonumber\\ & &+\frac{{d^\prime}^2 (24\cos^2(2\theta_0)+33)}{64}\Big(\frac{2}{3v_l^5}+\frac{1}{v_t^5}\Big)\Big],\nonumber
\end{eqnarray}
while for $\theta=\pi/4$ like in the experiment we have
\begin{eqnarray}
{T_1}^{-1}&=&\frac{(\hbar\omega_0)^3}{20\pi\rho\hbar^4}\Big(\frac{\hbar\omega_0}{\Delta}\Big)^2\Big[\frac{81{b^\prime}^2}{128}\Big(\frac{4}{3v_l^5}+\frac{2}{v_t^5}\Big)+\frac{33{d^\prime}^2}{64}\Big(\frac{2}{3v_l^5}+\frac{1}{v_t^5}\Big)\Big].
\label{T1s}
\end{eqnarray}
Taking ratio between Eqs.~(\ref{T1r}) and (\ref{T1s}) and substituting parameters other than $\hbar\omega_0$ and $\Delta$, we obtain the relation $\hbar\omega_0/2\Delta = 0.765\sqrt{T_{1,\text{Relaxed}}/T_{1,\text{Strained}}}$. 
Substituting $T_{1,\text{Relaxed}}=85~\mu\text{s}$ and $T_{1,\text{Strained}} = 5~\text{ms}$ to this equation, we obtain $\hbar\omega_0/2\Delta \approx \tfrac{1}{10}$ as shown in the main text.
We note that, while this estimation is made on the assumption $\hbar\omega_0/\Delta\ll 1$, well coincides with a result obtained by exact diagonalisation approach with an accuracy of $\lesssim 1~\%$.





\end{document}